\documentclass[a4paper,fleqn]{cas-sc}

\usepackage{natbib}
\usepackage{graphicx}
\usepackage{caption}
\usepackage{subfig}
\usepackage{cellspace} 
\usepackage[percent]{overpic}  
\usepackage{isomath}
\usepackage{amsmath}
\usepackage{changepage}
\usepackage{soul}
\usepackage{lineno}

\DeclareUnicodeCharacter{99999}{μ}

\def\tsc#1{\csdef{#1}{\textsc{\lowercase{#1}}\xspace}}
\tsc{WGM}
\tsc{QE}
\tsc{EP}
\tsc{PMS}
\tsc{BEC}
\tsc{DE}

\begin{document}
\let\WriteBookmarks\relax
\def\floatpagepagefraction{1}
\def\textpagefraction{.001}
\shorttitle{\ }
\shortauthors{Reyes-Guerrero et~al.}
\title [mode = title]{Information content on Venusian aerosols in VIRTIS-M infrared data} 
\author[1]{J. Reyes-Guerrero}[
                        orcid=0009-0003-8040-5496]
\cormark[1]
\ead{jaime.reyes@ehu.eus}

\affiliation[1]{organization={Escuela de Ingeniería de Bilbao, Euskal Herriko Unibertsitatea (EHU), Bilbao, Spain}}

\author[1]{S. Pérez-Hoyos}

\author[1]{I. Garate-Lopez}

\cortext[cor1]{Corresponding author}

\begin{abstract}
\noindent Venus presents a complex cloud structure with aerosol particles of different sizes located mainly between 48 km and 70 km in altitude. However, the number of free parameters that can describe such a cloud structure is usually overwhelming, and a number of simplified parameterizations are commonly assumed. In this work, we re-analyze the information content on aerosol vertical distribution provided by the nightside infrared data collected by Visible and InfraRed Thermal Imaging Spectrometer (VIRTIS) onboard Venus Express. Starting from \cite{haus_self-consistent_2013} aerosol vertical distribution description, we use the archNEMESIS radiative transfer code and retrieval suite together with the Bayesian inference tool Multinest to compute the evidence supporting different models based on alternative choices of model parameters. This study analyzes locations at three different regions: mid-latitudes, the so-called cold collar, and the South Polar Vortex. We use a data cube that covers all of these regions simultaneously and additional observations distributed throughout the Venus Express mission. We find that these observations provide significant information about the peak particle number density, base altitude and layer thickness of aerosol modes 2, 2' and 3 when individual parameters are retrieved - a more informative description than the scaling `mode factors' often used in similar retrievals. Including mode 1 in our models, however,  does not provide a significant increase of Bayesian evidence. The peak of mode 2 particle number density in mid-latitude locations is obtained at around 66 km, which is similar to the a priori model, whereas in the cold collar and South Polar Vortex locations, the peak is located at altitudes as low as 58 km. In the middle and lower clouds, particles of modes 2' and 3 present higher variability in altitude with respect to the a priori model. Our retrieved cloud and temperature profiles are in general agreement with previous studies, although we find higher cloud tops in the South Polar Vortex locations. This initial computing effort provides an optimal parameterization that will allow us to study in greater detail the instantaneous horizontal cloud and temperature structure from the entire VIRTIS-M infrared database, which we will discuss in a forthcoming paper.
\noindent 
\end{abstract}


\maketitle

\section{Introduction}
The atmosphere of Venus and its clouds have been studied for decades. Since the first space missions to arrive at Venus, the Veneras and Pioneer Venus series, it was confirmed that Venus has the most complex cloud structure among the terrestrial planets of our solar system \citep{esposito_1983}. The main cloud deck extends approximately from 48 km up to 70 km with upper and lower hazes above and below these limits \citep{esposito_1997}. In situ data from the Particle Size Spectrometer onboard the Pioneer Venus Large Probe \citep{knollenberg_hunten} revealed different aerosol population modes that suggest a subdivision of the main cloud deck into upper, middle and lower clouds in terms of their microphysical properties.

The upper cloud presents submicron (mode 1) and micron sized (mode 2) particles and extends from about 56 to 65-70 km, with an upper boundary that varies latitudinally \citep{ragent_particulate_1985}. While the upper cloud is easy to observe in a broad spectral range from UV to thermal IR, the middle and lower clouds, located between approximately 48 and 56 km, are mainly sensed through the IR windows between 1.5 and 2.5 $\mathrm{\mu}$m \citep{allen_cloud_1984}. These layers contain mode 1 and 2 particles but also contain particles with sizes even higher than 3 $\mathrm{\mu}$m, called modes 2' and 3. Regarding the composition of these aerosol modes, ground based observations of reflected sunlight linear polarization were best explained by a solution of sulfuric acid \citep{hansen_hovenier}. This was later confirmed with data from Venera 15 and 16 orbiters and entry probes and balloons of Vega 1 and 2 \citep{zasova_structure_2007}. Sulfuric acid concentration has also been studied through ground observations \citep{arney_spatially_2014} and radiative transfer modeling \citep{barstow_2012}. However, the nature of aerosol particles is still under discussion, especially for the lower cloud, for which a heterogeneous composition containing water, ferric sulfate and sulfuric acid has been proposed in a recent reanalysis of Pioneer Venus data \citep{moguletal2025}.

Understanding the cloud structure of Venus is key to comprehending different aspects of the entire atmosphere. At the upper cloud altitude levels, sulfuric acid is photochemically produced from SO$_2$ and H$_2$O \citep{titov_clouds_2018} and
the temperature structure shows strong variability \citep{zasova_structure_2006}.
With respect to atmospheric dynamics, nightside images at 1.74 $\mathrm{\mu}$m, that probe the structure of the lower cloud, constitute the main data to study lower cloud dynamics \citep{hueso_assessing_2012,hueso_six_2015} and images at 3.8 and 5.1 $\mathrm{\mu}$m are used to measure winds in the polar upper clouds (\citeauthor{peralta_solar_2012}, \citeyear{peralta_solar_2012}; \citeauthor{garate_lopez_2013}, \citeyear{garate_lopez_2013}). Cloud tracking from  ultraviolet and visible observations allows us to measure winds in the upper cloud of mid-latitude region \citep{peralta_reanalysis_2007,khatuntsev_cloud_2013}, where zonal wind speed reaches its maximum \citep{sanchez-lavega_atmospheric_2017}. The radiative energy balance is also affected by the cloud structure. \cite{lee_radiative_2015} studied the radiative forcing variation due to clouds in the mesosphere and demonstrated that cloud top altitude affects both cooling and heating rates. \cite{garate-lopez_latitudinal_2018} reached the same conclusion when analyzing the effects of latitudinal cloud variation in the Venus Planetary Climate Model (V-PCM). Analyses of the global energy budget of the planet (e.g, \cite{haus_radiative_2015,haus_radiative_2016}) also showed the influence of the cloud structure.

With a payload mainly inherited from Mars Express and Rosetta missions, Venus Express was the first mission to Venus for the European Space Agency. Its Visible and Infrared Thermal Imaging Spectrometer (VIRTIS) was designed to analyze both Venus atmosphere and surface through its three channels \citep{piccioni_virtis_2007}. Data from this instrument have been extensively used to study different aspects of the atmosphere such as minor gases species \citep{tsang_correlated-k_2008,tsang_correlations_2010,bezard_water_2009,haus_lower_2015} as well as atmospheric dynamics \citep{hueso_assessing_2012,hueso_six_2015,garate_lopez_2013, garate-lopez_potential_2016} and, of course, the vertical structure of the clouds \citep{wilson_evidence_2008,ignatiev_altimetry_2009,barstow_2012,lee_vertical_2012,haus_self-consistent_2013,haus_atmospheric_2014,magurno_retrieval_2017}. \cite{haus_self-consistent_2013} presented one of the most detailed parameterizations for the aerosols vertical distribution that was used in their following work \citep{haus_atmospheric_2014} to analyze the average thermal and cloud structure of both hemispheres using VIRTIS infrared measurements.

Atmospheric retrievals through radiative transfer calculations are an inverse multiparametic problem, as the effect of different parameters on the simulated radiance is coupled and many alternative sets of parameters provide satisfactory fits to the data. This is the case of Venus infrared windows, from which it is not easy to disentangle the effect of temperature, cloud and minor gases structure on the detected radiance \citep{tsang_correlations_2010,haus_radiative_2010, garcia_munoz_model_2013}. Regarding the aerosol vertical distribution, there is no previous exhaustive analysis on the number of free parameters required and supported by the observations. Bayesian inference techniques such as MultiNest \citep{feroz_multinest_2009, buchner_x-ray_2014} sample the entire free parameter space to calculate their posterior probability distribution functions and the Bayesian evidence of the model, a value that evaluates the probability the model represents the observations regardless of the value of its parameters. This allows the comparison of models with different parameterizations that could satisfactorily fit the data, and thus determine which one is better supported by the observations and hence the preferred model. 

The goal of this paper is to re-analyze Venus' infrared spectra using Bayesian inference techniques to maximize the information that can be obtained on the vertical distribution of different aerosol modes, which will be extremely useful, for instance, when extracting data from future missions such as EnVision. In addition, we seek to define a suitable parameter set, based on \cite{haus_self-consistent_2013} model, to study the horizontal distribution of clouds and its possible temporal evolution from the entire VIRTIS database in a forthcoming paper.

This work is organized as follows. In Section \ref{sec:Data}, we briefly describe VIRTIS measurements, and present the observations selected for this paper as well as the preprocessing pipeline applied to the data. In Section \ref{sec:Methodology}, we describe the a priori atmospheric model, the used radiative transfer code and the methodology applied. The results are presented in Section \ref{sec:Results} and further discussions are included in Section \ref{sec:Discussion}. Finally, our main conclusions are summarized in Section \ref{sec:Conclusions}.

\section{Data}\label{sec:Data}
\subsection{VIRTIS}
The Visible and Infrared Thermal Imaging Spectrometer (VIRTIS) onboard Venus Express mission operated through three channels: VIRTIS-M-VIS (imaging spectrometer in the 0.3–1 $\mathrm{\mu}$m range), VIRTIS-M-IR (imaging spectrometer in the 1–5 $\mathrm{\mu}$m range) and VIRTIS-H (aperture high-resolution spectrometer in the 2–5 $\mathrm{\mu}$m range). All channels were able to work simultaneously to provide combined observations of the same area with a field of view of 64x64 mrad$^2$ for the VIRTIS-M channels and 0.58x1.75 mrad$^2$ (IFOV) for the VIRTIS-H channel \citep{drossart_scientific_2007}. Due to the elliptical polar orbit of the mission, VIRTIS-M was able to perform spectral mapping of several regions in the southern hemisphere in off-pericenter observations, and full-disk maps can be constructed in 3x3 mosaic images from apocenter observations at a distance of 66,000 km \citep{piccioni_virtis_2007}.

Both VIRTIS-M-vis and VIRTIS-M-IR data are stored in hyperspectral images called `data cubes'. The maximum dimensions of these data cubes at full FOV correspond to 256x256 spatial pixels by 432 wavelengths with a spectral resolution of $\sim$15 nm. Therefore, each band of this array contains an image at a specific wavelength while, at the same time, a spectrum can be obtained from each pixel. Time exposures in VIRTIS measurements varies between 0.02 and 18 s. However, to avoid saturation at wavelengths longer than 4.2 $\mathrm{\mu}$m, a typical value of 1.0 s was used \citep{piccioni_virtis_2007}.
\subsection{Data selection}
Between 2006 and 2008, Venus Express completed more than 900 orbits taking several VIRTIS-M-IR measurements per orbit, resulting in more than 4,500 data cubes covering different regions of both hemispheres, but mainly the southern one. We aim at studying the latitudinal and time variability of the atmosphere of Venus in terms of thermal and cloud structure. However, before analyzing the entire VIRTIS-M-IR data base, it is mandatory to know how to extract as much information as possible from these  measurements regarding the aerosol vertical distribution. In this work, we focus on this issue, from a careful selection of observations, leaving the analysis of the VIRTIS database for a forthcoming publication.

Above 70 km altitude, the temperature increases from equator to pole \citep{tellmann_structure_2009} but, at $\sim$62 km altitude and between 60$^\circ$ and 80$^\circ$, in both hemispheres, the temperature field shows thermal inversions in a region called the `cold collar' \citep{taylor_structure_1980}. In this region, a collar of cold air circles each pole and upwelling motions can occur, as predicted by GCMs \citep{garate-lopez_latitudinal_2018}. Moreover, the cold collar encloses a warm vortex that is, on average, $\sim$30 K warmer than the surroundings  \citep{garate-lopez_instantaneous_2015}. On the other hand, the latitudinal variation of the cloud tops has been extensively studied (\citeauthor{lee_vertical_2012}, \citeyear{lee_vertical_2012};  \citeauthor{haus_self-consistent_2013}, \citeyear{haus_self-consistent_2013}, \citeyear{haus_atmospheric_2014};  \citeauthor{grassi_venus_2014}, \citeyear{grassi_venus_2014}) with a general agreement on a poleward decrease of about 4-10 km and long-term variations of about 3 km \citep{ignatiev_altimetry_2009}. To appropriately represent this spatial variability on temperature and cloud top altitude, three regions will be analyzed in this work: mid-latitude, sub-polar (cold collar) and polar (South Polar Vortex) regions.

Thermal emission from the mesosphere is detected in VIRTIS-M-IR spectra above 3.5 $\mathrm{\mu}$m \citep{haus_radiative_2010}, so we must choose data cubes with a short exposure time to avoid saturation in this range. Mid and sub-polar latitudes do not usually show any structure at these wavelengths, but in all three selected regions bright and dark features come into view at other emission bands, such as 1.74 and 2.3 $\mathrm{\mu}$m, where radiation originates at lower levels in the atmosphere and is attenuated by the deep clouds \citep{crisp_nature_1989}. To take this into account, a spectrum under `bright' and `dark' conditions from each region was extracted by visually inspecting the band centered at 2.3 $\mathrm{\mu}$m.

\begin{figure}
\centering
  \includegraphics[width=0.3\textwidth]{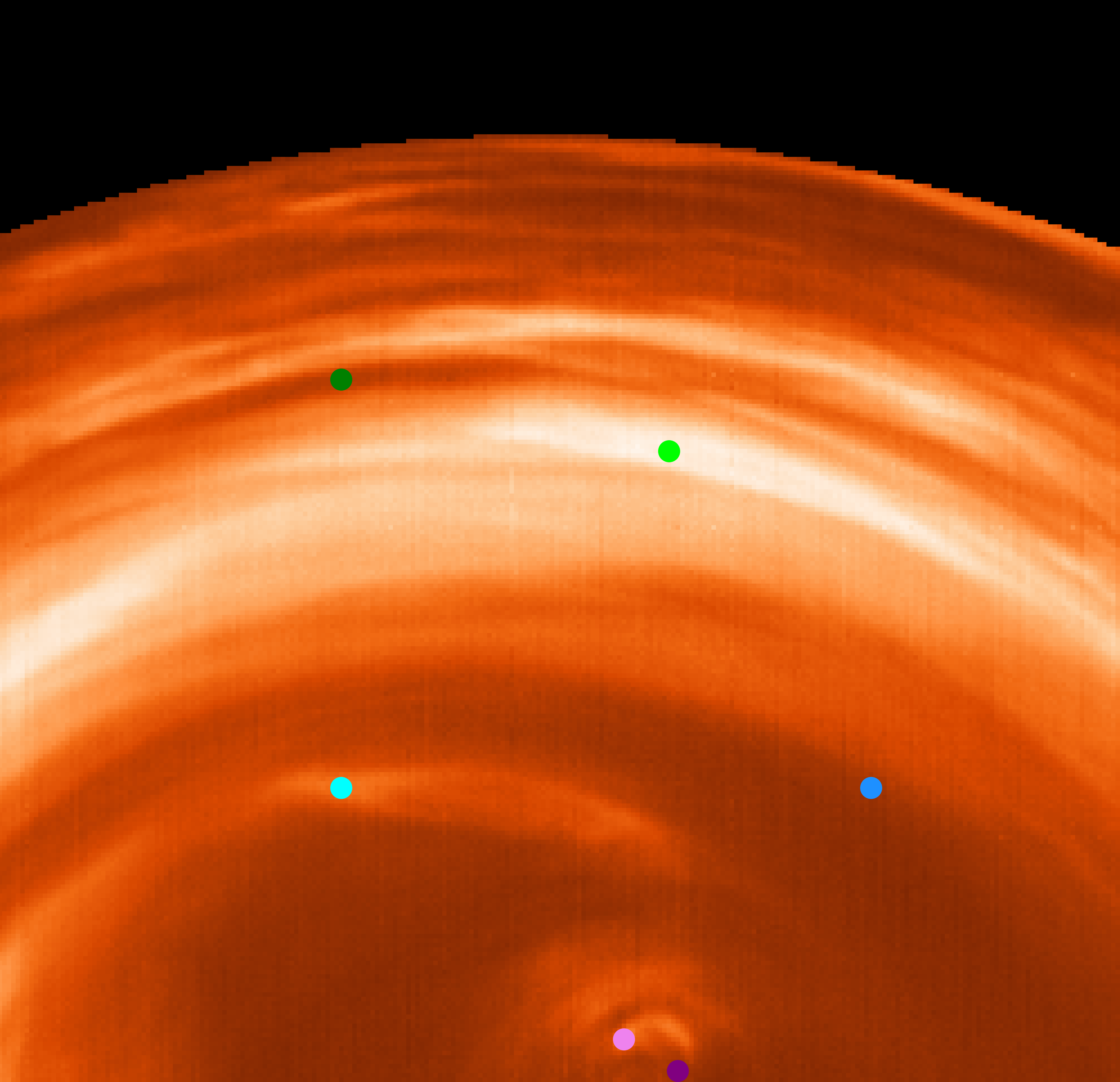}
  \label{fig:sub1}
  \includegraphics[width=0.3\textwidth]{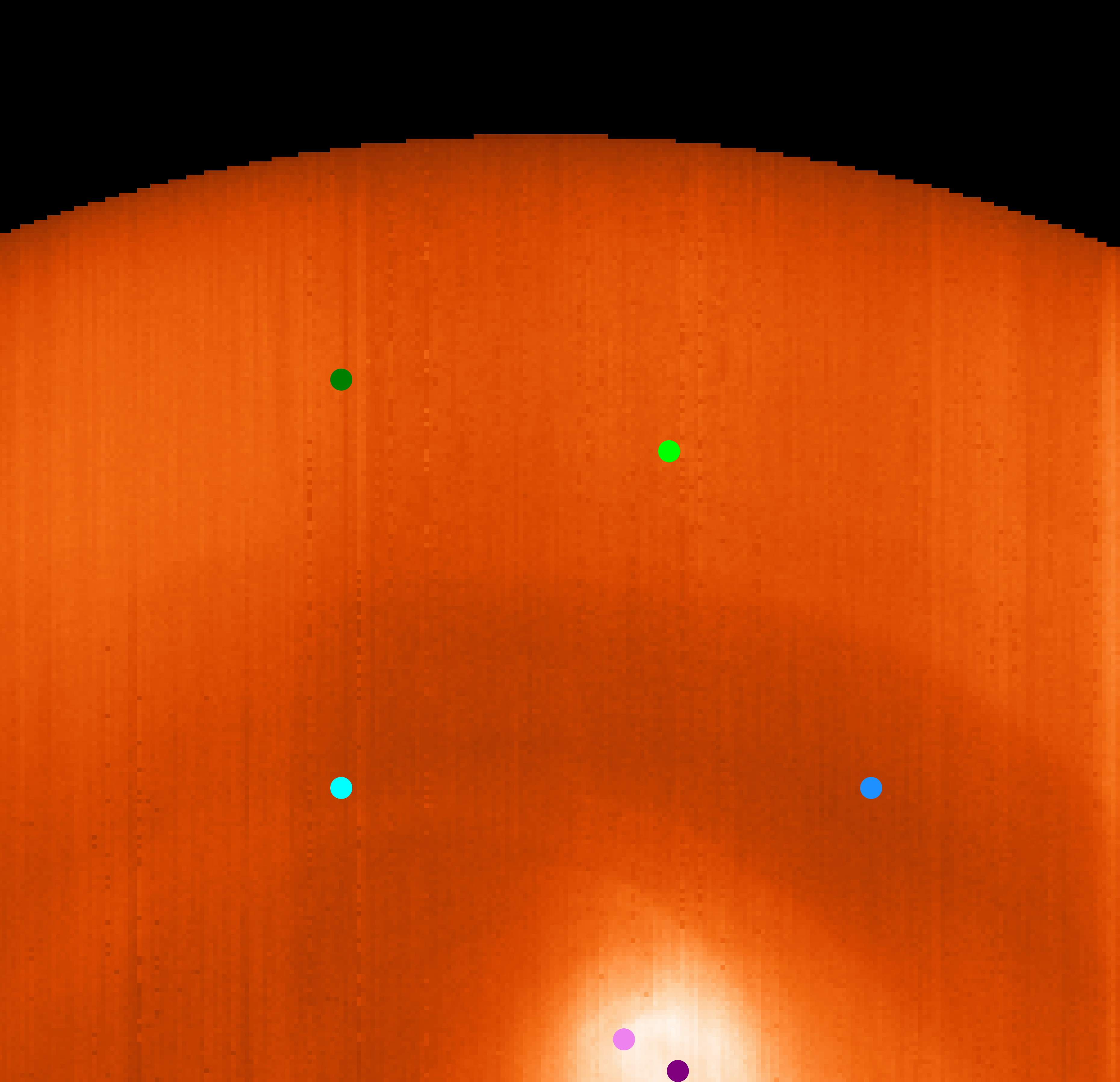}
  \label{fig:sub2}
\caption{Selected locations of data cube VI0025\_07: mid-latitude (green), cold collar (blue) and South Polar Vortex (purple). Lighter dots indicate `bright' condition and darker dots indicate `dark' condition. Left panel shows band 135 ($\sim 2.3 $ $\mathrm{\mu}$m) and right shows band 420 ($\sim 5$ $ \mathrm{\mu}$m).} 
\label{fig:locations}
\end{figure}

The data cube VI0025\_07, i.e, measurement 7 of orbit 25, is well suited for our study since it shows mid-latitudes, the cold collar and the South Polar Vortex at the same time with high spatial resolution and appropriate exposure time. The selected locations to be studied are shown in Figure \ref{fig:locations}. To have a temporally distributed sample of observations, we will also analyze similar locations from six additional data cubes distributed throughout the mission time. A summary of the observations, including location over the image plane and the planetary disk, as well as the emission angle value, can be found in Table \ref{table:observations}.
    
\begin{table}
\begin{tabular}{cccccccc}
\toprule
Data cube                   & Date                        & Region (location)                      & Sample & Line  & Latitude & Longitude & Em. angle\\ \midrule
\multirow{6}{*}{VI0025\_07} & \multirow{6}{*}{15/05/2006} & Mid-latitude (bright)       & 155    & 140 & -54.47$^\circ$ & 350.25$^\circ$ & 52.92$^\circ$ \\
                            &                             &  Mid-latitude (dark)        & 80     & 155  & -47.97$^\circ$ & 8.24$^\circ$ & 58.07$^\circ$\\
                            &                             & Cold collar (bright)        & 82     & 65   & -67.45$^\circ$ & 22.51$^\circ$ & 36.00$^\circ$\\
                            &                             & Cold collar (dark)        & 200    & 65   & -69.09$^\circ$ & 330.63$^\circ$ & 37.35$^\circ$ \\
                            &                             & South Polar Vortex (bright) & 145    & 9   &-80.76$^\circ$ & 1.76$^\circ$ & 24.29$^\circ$ \\
                            &                             & South Polar Vortex (dark)  & 157    & 2  & -82.06$^\circ$ & 349.83$^\circ$ & 23.40$^\circ$\\ \midrule
VI0038\_00                  & 28/05/2006                  & South Polar Vortex (bright) & 180     & 80  & -77.59$^\circ$ & 9.10$^\circ$ & 33.68$^\circ$ \\
VI0095\_07                  & 24/07/2006                  & Cold collar (dark)          & 55     & 10  & -66.55$^\circ$ & 171.54$^\circ$ & 35.25$^\circ$\\
VI0310\_00                  & 24/02/2007                  & South Polar Vortex (dark)   & 140    & 155 & -78.63$^\circ$ & 128.69$^\circ$ & 20.38$^\circ$ \\
VI0411\_03                  & 06/06/2007                  & Mid-latitude (dark)         & 65     & 35   & -33.74$^\circ$ & 137.02$^\circ$  & 51.00$^\circ$ \\
VI0726\_02                  & 15/04/2008                  & Cold collar (bright)         & 100    & 105  & -64.02$^\circ$ & 334.58$^\circ$ & 30.50$^\circ$\\
VI0818\_11                  & 17/07/2008                  & Mid-latitude (bright)       & 85     & 45   &-41.75$^\circ$  & 294.63$^\circ$ & 38.78$^\circ$ \\ \bottomrule
\end{tabular}
\caption{Summary of observations used in this work.}
\label{table:observations}
\end{table}

\subsection{Preprocessing}
As for any instrument, VIRTIS data needed to be calibrated for the analysis of measurements through an entire pipeline described in \cite{cardesin_moinelo_calibration_2010}. However, nightside data are still affected by interfering straylight that impacts the measured infrared radiance values, especially at wavelengths shortward of 3 $\mathrm{\mu}$m. \cite{kappel_refinements_2012} put great effort into developing new preprocessing techniques to exploit the limits of what can be achieved with these data and also made a description of the different possible sources of straylight.

Many straylight correction methods (\citeauthor{mueller_venus_2008}, \citeyear{mueller_venus_2008}, \citeyear{mueller_derivation_2017}, \citeyear{mueller_multispectral_2020}; \citeauthor{kappel_refinements_2012}, \citeyear{kappel_refinements_2012}; \citeauthor{mcgouldrick_discovery_2017}, \citeyear{mcgouldrick_discovery_2017}) are based on the selection of several wavelengths where Venus infrared spectrum is expected to be dark and on the assumption that any light seen at those wavelengths must be contamination from other sources. Radiance measurements at these wavelengths, hereafter `dark bands', can provide an estimation of such contamination, which includes, but is not limited to, straylight. In the following, we will make no assumption on its origin, although we will accept that it will depend on both wavelength and observation geometry.

Our correction approach works as follows. To reduce the effect of twilight, where light from the dayside can be scattered to the nightside, measurements are restricted to a solar angle higher than 98º. We chose as dark bands a combination of those proposed by \cite{kappel_refinements_2012} and \cite{mueller_derivation_2017}, i.e, bands 4, 22, 34-39, and 91-94 ($\sim$1.05, 1.22, 1.34-1.39, and 1.88-1.89 $\mathrm{\mu}$m, respectively), and, for each of them, a grade 2 polynomical fit between phase angle and radiance is performed to consider the geometric dependence. Finally, this modeled interfering radiance at dark bands is linearly extrapolated to the range 1-3 $\mathrm{\mu}$m to consider the spectral dependence. Note that this correction assumes no spectral features in the straylight spectral shape, which is approximated linearly between dark bands. The correction is done by subtracting this model from the uncorrected VIRTIS spectra at each pixel of a data cube. A correction of the odd-even effect was also applied using the correction kernel proposed by \cite{kappel_refinements_2012}.

VIRTIS data do not include an estimation of the radiance uncertainty at each wavelength.  \cite{cardesin_moinelo_calibration_2010} estimates a typical value of 15\%-20\% of the absolute radiance and highlights that this type of uncertainty is very difficult to estimate for planetary remote sensing instruments, since there is no ground thruth available for reference. It is also a frequent approach when using VIRTIS spectra to perform a spatial binning  \citep{lee_vertical_2012,garate-lopez_instantaneous_2015,magurno_retrieval_2017} which increases signal-to-noise ratio (SNR) and allows to use the Noise Equivalent Spectral Radiance (NESR) divided by the number of points averaged as an estimation of the error \citep{grassi_retrieval_2008,irwin_spatial_2008}. However, in this paper, we used spectra at isolated pixels, and hence we had to estimate the data uncertainty considering possible errors introduced by the straylight and odd even corrections. \cite{kappel_refinements_2012} suggests that straylight can be neglected beyond 3 $\mathrm{\mu}$m, a spectral region where the SNR also increases \citep{piccioni_virtis_2007}. Therefore, we decided to apply different uncertainty values for the ranges 1.7-3 $\mathrm{\mu}$m and 3-5.2 $\mathrm{\mu}$m. An uncertainty value of 15\% of the measured radiance is applied in the first range and of 5\% in the second. Moreover, to ensure that the radiance uncertainty does not achieve an extremely low value, especially at locations under `dark' conditions, we set a minimum value of the uncertainty of $1.5\times 10^{-3}$ Wm$^{-2}$sr$^{-1}$$\mathrm{\mu}$m$^{-1}$ for both spectral ranges. These applied uncertainties allow us to achieve satisfactory fits to the measurements and to establish an appropriate criterion on the fit quality, as described later in Sec. \ref{sec:Results}.

\section{Methods} \label{sec:Methodology}
\subsection{Atmospheric model} \label{sec:atm_model}
\begin{figure}
    \centering
    \includegraphics[width=0.5\linewidth]{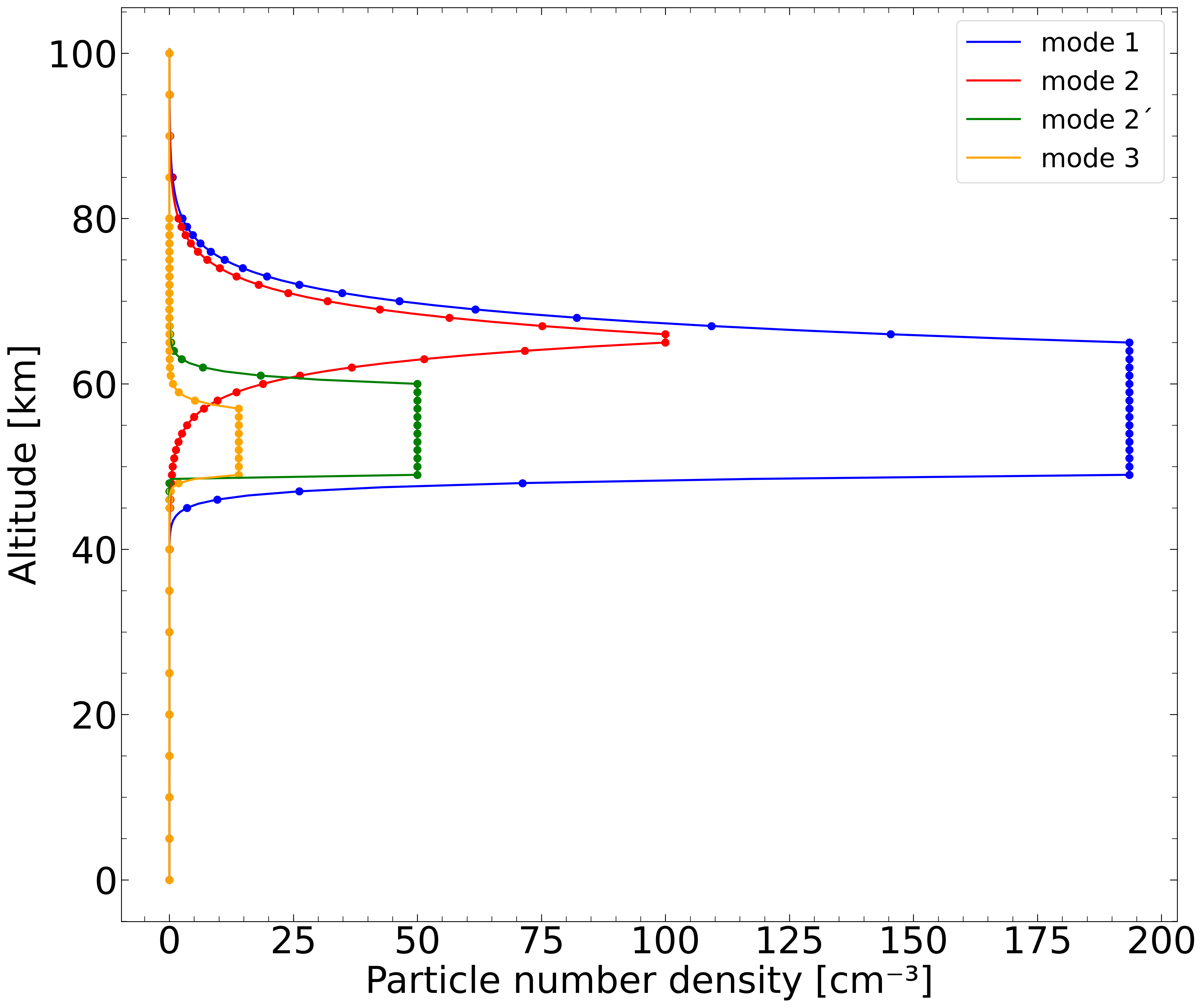}
    \caption{Aerosol vertical distribution of the standard cloud model from \cite{haus_radiative_2016}. Dots represent the evaluation of each mode at the altitude grid of our model.}
    \label{fig:haus}
\end{figure}

The a priori atmospheric model follows the particle altitude distribution proposed by \cite{haus_self-consistent_2013} (Figure \ref{fig:haus}). Their description is very convenient since each of the four aerosol modes is described by the same analytical expression with five parameters (Eq. \ref{eq:haus}): particle number density ($N_0$) at a specific base altitude ($z_b$), layer thickness of constant peak particle number density ($z_c$) and upper ($H_{up}$) and lower ($H_{lo}$) scale heights. This description is flexible enough to picture many different aerosol vertical distributions with a reasonable number of free parameters, so we decided to take it as our prior cloud distribution. The values for each parameter are taken from the latest version of this model, proposed in \cite{haus_radiative_2016} and listed in Table \ref{table:haus}.

\begin{table}
\begin{tabular}{cccccc}
\toprule
Mode & ${N_0}$ {[}cm$^{-3}${]} & $z_b$ {[}km{]} & ${z_c}$ {[}km{]} & ${H_{up}}$ {[}km{]} & \multicolumn{1}{l}{${H_{lo}}$ {[}km{]}} \\ \midrule
1             & 193.5                          & 49.0                      & 16.0                      & 3.5                        & 1.0                                            \\ 
2             & 100.0                            & 65.0                      & 1.0                     & 3.5                        & 3.0                                           \\ 
\ 2'            & 50.0                             & 49.0                    & 11.0                    & 1.0                        & 0.1                                            \\ 
3             & 14.0                             & 49.0                    & 8.0                     & 1.0                        & 0.5                                            \\ \bottomrule
\end{tabular}
\caption{Parameter values of the standard cloud model from \cite{haus_radiative_2016}.}
\label{table:haus}
\end{table}

All aerosol modes are assumed to be composed of spherical H$_2$SO$_4$ droplets concentrated at 75\% with refractive indices data taken from \cite{palmer_optical_1975}. For size distributions, we followed values proposed by \cite{pollack_near-infrared_1993}, i.e., log-normal size distributions with modal radii of 0.3, 1.0, 1.4, 3.65 $\mathrm{\mu}$m,  and standard deviations of 0.44, 0.25, 0.21 and 0.25 $\mathrm{\mu}$m. Temperature profiles are constructed from latitude-dependent VIRA-2 \citep{zasova_structure_2006} profiles above 50 km and, below 40 km, they are constructed from VIRA-1 profiles \citep{seiff_models_1985}, which are latitudinally constant below 32 km. Both datasets are connected between 40 and 50 km with a linear interpolation.
\begin{equation}
    N(z) = 
    \begin{cases}
        N_0(z_b)exp[-(z-(z_b+z_c)/H_{up}],\ z>(z_b+z_c)\\
        N_0(z_b), \ (z_b+z_c)\geq z \geq z_b \\
        N_0(z_b) exp[-(z_b-z)/H_{lo}], \ z<z_b
    \end{cases}
    \label{eq:haus}
\end{equation}
Absorption cross-sections of the gases are taken from precomputed k-tables calculated with the following parameters. For CO$_2$, the k-table was generated using HITEMP 2010 dataset \cite{rothman_hitemp_2010}, considering self-broadening, a sub-Lorentzian line shape taken from \cite{tonkov_measurements_1996} and a line cutoff at 120 cm$^{-1}$. We are aware that a new HITEMP CO2 line database has recently been released \citep{hargreaves_updating_2025} that improves accuracy, as well as spectral and isotopologue coverage. However, after some tests, we found that HITEMP2010 database better reproduced the shape of the 2.3 $\mathrm{\mu}$m window and the wings of the 4.3 $\mathrm{\mu}$m CO$_{2}$ band with the cut-off value used here, similarly to previous releases. These minor discrepancies are beyond the scope of this paper, but may merit future research. For all other gas species, the HITRAN 2020 dataset \citep{gordon_hitran2020_2022} is used and air-broadening is considered. CO$_2$ continuum absorptions are also included with values of (in cm$^{-1}$ amagat$^{-2}$): $5 \times 10^{-9}$ for 1.74 $\mathrm{\mu}$m band \citep{de_bergh_water_1995} and $40 \times 10^{-9}$ for 2.3 $\mathrm{\mu}$m band \citep{de_bergh_water_1995,tonkov_measurements_1996,kappel_refinements_2012}. Note that these CO$_2$ continuum absorptions take into account not only collision-induced absorptions but also dimer absorptions and far wings of allowed bands, as described by \cite{snels_carbon_2014}. In our model, the atmosphere is divided into 49 layers with a vertical resolution of 5 km between 0-45 km and 80-100 km and 1 km between 45-80 km.

\subsection{Radiative transfer code} \label{sec:rt_code}
The radiative transfer and retrieval suite used in this work is archNEMESIS \citep{alday_archnemesis_2025}, a recently released version of NEMESIS \citep{irwin_nemesis_2008} adapted to Python. Several previous works have already used NEMESIS to simulate Venus' spectra in both visible \citep{perez-hoyos_venus_2018} and infrared ranges (\citeauthor{tsang_correlated-k_2008}, \citeyear{tsang_correlations_2010}; \citeauthor{barstow_2012}, \citeyear{barstow_2012}). This code was based on an optimal estimator scheme \citep{rodgers_2000} to perform the retrieval of atmospheric properties. This scheme is intended to minimize the difference between observed and simulated spectra, using the same geometrical configuration, starting from a priori information of parameter values and uncertainties and performing ideally the minimum number of forward model evaluations.

However, a new feature of archNEMESIS is the implementation of MultiNest (\citeauthor{feroz_multinest_2009}, \citeyear{feroz_multinest_2009}; \citeauthor{buchner_x-ray_2014}, \citeyear{buchner_x-ray_2014}), an inference tool based on Nested Sampling Monte Carlo algorithms \citep{skilling_nested_2006}. 
These algorithms also require parameter a priori information, but this is given as probability distributions of their values, from which the posterior probability distribution is computed. The algorithm is designed so that a large enough number of iterations erases all information from the prior, and thus the whole free parameter space is scanned in an optimal way. In each of these iterations, a given atmospheric scheme is evaluated for a combination of input parameter values, and a likelihood value is associated to it after comparing observed and simulated spectra. The likelihood value can be computed as $\mathcal{L}\propto exp(-\chi^2/2)$ \citep{trotta_bayes_2008}, where $\chi^2$ is
the quadratic deviation between modeled and observed spectra defined as $\chi^2=\sum_{1}^{n} ((m_i-s_i)/e_i)^2$. In this definition, $m_i$ are the measured radiances of estimated error $e_i$, $s_i$ are the simulated radiances and $n$ is the number of spectral points. This metric is widely used to estimate the precision to which measured spectra can be fitted \citep{irwin_spatial_2008,grassi_venus_2014,perez-hoyos_venus_2018} and using it in the likelihood value calculation allows us to consider an evaluation that achieves a good fit to the data more likely to represent the actual status of the atmosphere. 

Nested sampling techniques are able to compute the Bayesian evidence of the atmospheric model, considering all the definite schemes evaluated. This value is just the average of the likelihood under the prior for a specific model choice (see Section 4.2 of \citealp{trotta_bayes_2008}). A model in which a large number of evaluations achieve a high likelihood value (i.e., a good fit to the data) will have a higher Bayesian evidence value. On the other hand, models in which only a few combinations of input parameters lead to a good fit will have a lower Bayesian evidence value. Therefore, the existence of a single combination of atmospheric parameters that achieves an excellent fit to the data does not necessarily mean that the Bayesian evidence of its underlying model will be highly rated. It can be stated then that the Bayesian evidence gives information about the probability that the entire model statistically represents the observations. Moreover, this allows us to perform comparisons between different models using the Bayes factor, which is defined as $B_{12}=\ln Z_1 - \ln Z_2$, where $Z_1$ and $Z_2$ represent the Bayesian evidence of competing models 1 and 2. In the atmospheric properties retrieval problem, it is quite frequent to deal with different sets of free parameters that provide similar fits to the data, and precisely the model comparison via the Bayes factor can shed light on the issue of which one is better or more probable. The Bayes factor not only depends on the likelihood, but it also measures whether the use of a more complex model (which can be roughly estimated by the number of free parameters) is sustained by the data (see Section 4.4 of \citealp{trotta_bayes_2008}). The model comparison using the Bayes factor is usually done following the Jeffrey's scale \citep{jeffreys1998theory,trotta_bayes_2008}, shown in Table \ref{table:jeffreys}. When the Bayes factor $B_{12}$ is greater than 5, there is a strong evidence to choose model 1 over model 2. In case $B_{12}$ is lower than 1, there is not enough evidence to favour one of them, and the one with fewer free parameters is usually chosen, following the Occam's razor principle. In this work, we will settle for a difference of 1 in the Bayes factor to select a model, even though this evidence is considered as `weak' on Jeffrey's scale.

As other algorithms \citep{metropolis_1953,hastings,gibbs_1984}, MultiNest also calculates the posterior probability distribution for each free parameter, allowing the parameter determination as well as the visualization of possible correlation between parameters by representing the marginalized probability distribution in a corner plot. Hence, both model selection and parameter estimation are conveniently computed, at the cost of performing enough evaluations of the forward model. As an example, a typical retrieval in this work, with 20 free parameters, requires more than 12,000 evaluations, which usually demands parallelization to be feasible on reasonable time scales. To speed up the process, we reduce the spectral resolution to 49 spectral points. This is obviously much slower than an optimal estimator retrieval, which may require only a few tens of evaluation, but the information provided by the Bayesian inference algorithms is much more complete. 
\begin{table}
\begin{tabular}{cc}
\toprule
${B = \ln Z_1 - \ln Z_2}$   & Evidence \\ \midrule
\textless{}1.0    & Inconclusive      \\ 
1.0 - 2.5                            & Weak              \\ 
2.5 - 5.0                          & Moderate          \\ 
\textgreater{}5.0 & Strong            \\ \bottomrule
\end{tabular}
\caption{Jeffrey's scale \citep{jeffreys1998theory,trotta_bayes_2008}.}
\label{table:jeffreys}
\end{table}

\subsection{Retrieval strategy} \label{sec:retrievals}
Studying the information that VIRTIS-M infrared data provides about aerosols vertical distribution can be a daunting task, since there are too many aspects that influence the detected radiance; each aerosol mode is defined by several parameters and the temperature vertical profile also plays a key role. For this reason, we have to define a strategy to correctly perform the comparison of models with different parameterizations. Following strategies previously used with Bayesian inference techniques in the context of exoplanets atmospheres \citep{lueber_information_2024, roy-perez_role_2025}, we ran sets of retrievals to first explore the information content on each aerosol mode and then the influence of the parameterization used for the aerosol modes on the retrieved aerosol vertical distribution. 

\begin{figure}
    \centering
    \includegraphics[width=1\linewidth]{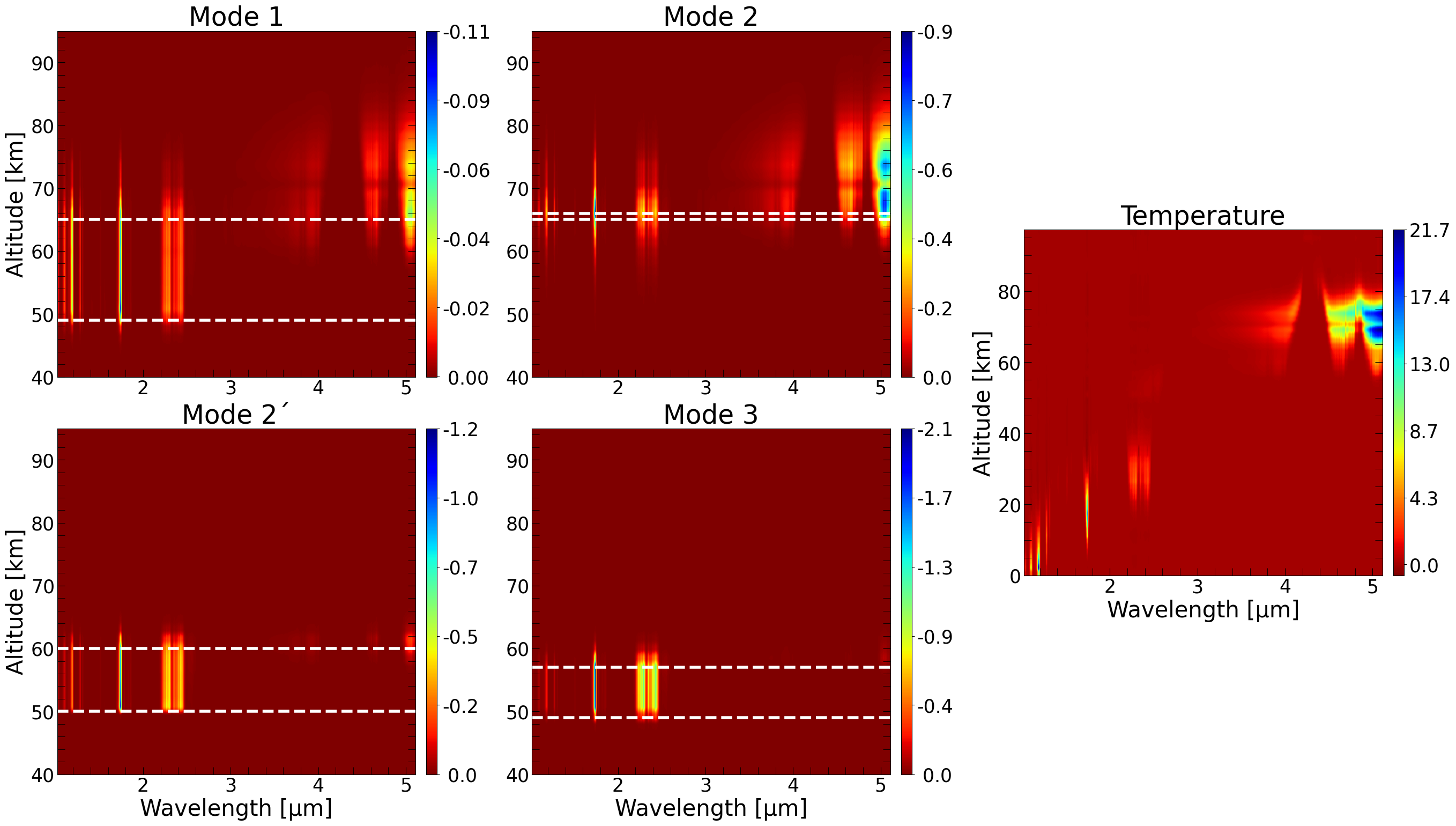}
    \caption{Jacobian matrices of the four aerosol modes and temperature. Colorbar values indicate the relative radiance difference (\%) between the spectra of the models with modified and a priori aerosol abundance or temperature. Horizontal dashed white lines show a priori base altitude ($z_b$) and altitude up to which the particle number density is constant ($z_b$+$z_c$) for each aerosol mode.}
    \label{fig:jacobians}
\end{figure}

We started inspecting the sensitivity of the atmospheric model to each aerosol mode and to the temperature profile. A first approach to this is to calculate their Jacobian matrices, i.e., partial derivatives of the modeled radiance with respect to a free parameter, as a function of wavelength and altitudes (Figure \ref{fig:jacobians}). These were obtained by varying the abundance of each aerosol mode (or temperature) by 1\% at each altitude level and calculating the relative radiance with respect to the spectrum obtained with a priori aerosol abundances and temperature profile. Mode 1 particles produce the weakest differences in the spectrum, whereas the rest of aerosol modes produce similar radiance differences, although mode 3 produces overall higher radiance differences. A priori cloud base and top of constant number density for each aerosol mode are also shown in Figure \ref{fig:jacobians} to help with the sensitivity analysis. In the 3.5-5 $\mathrm{\mu}$m range, where the cloud top is probed, the atmospheric model is clearly more sensitive to modes 1 and 2, although modes 2' and 3 cloud top layer also show little but not negligible contribution. Modes 1 and 2 are detectable at high atmospheric levels, well above their constant particle number density ranges, because of their large upper height scales. However, below 2.5 $\mathrm{\mu}$m, the upwelling radiation from layers below the clouds makes the four models detectable but strongly coupled, so their individual contributions are difficult to infer from this range alone. On the other hand, the highest values were obtained for temperature variations. This shows that the thermal and cloud contributions to the radiance are highly coupled and therefore should be retrieved together, as suggested by \cite{garcia_munoz_model_2013}. Moreover, surface properties, i.e., surface temperature, emissivity and altitude, do also contribute to the detected nightside spectrum, reaching a contribution of 96.4 \% in the first NIR window at 1.02 $\mathrm{\mu}$m \citep{haus_radiative_2010, mueller_multispectral_2020}. In this work we focused on the retrieval of aerosol vertical distribution and, below 1.5 $\mathrm{\mu}$m, the single scattering albedo of all the aerosol modes are near unity, and hence it is not easy to disentangle their individual contribution. Therefore, we decided to constrain the spectral range to 1.7-5.2 $\mathrm{\mu}$m.

While it is possible to retrieve the temperature at a number of altitude levels (potentially at all of them) with MultiNest, this would increase heavily the number of free parameters and hence scale exponentially the computation time. The common solution is to parameterize the thermal profile with a few free parameters but this would erase some features in the temperature field that are also interesting. The atmospheric locations studied in this work were selected due to their thermal and cloud variability, as already discussed, and therefore it is not suitable to assume the same thermal structure for each of them. For these reasons, we decided to retrieve first the temperature vertical profile for each location to be used as a priori profile in MultiNest retrievals. These burn-in retrievals were performed using the optimal estimator scheme of archNEMESIS. Following the approach of previous studies focused on retrieving mesospheric temperature structures \citep{lee_vertical_2012,grassi_venus_2014,haus_atmospheric_2014,garate-lopez_instantaneous_2015}, the temperature profile between 55-80 km was retrieved together with modes 1 and 2 vertical distributions using the spectral range 3.5-5 $\mathrm{\mu}$m. Modes 2' and 3 vertical distributions, on the other hand, were retrieved using the transparency windows at 1.7 and 2.3 $\mathrm{\mu}$m. At these wavelengths, carbon dioxide continuum absorption has a large effect and, since there are not many reliable laboratory measurements of this opacity at Venus deep atmosphere conditions \citep{snels_carbon_2014}, it is frequent to consider CO$_2$ continuum opacity as a correction factor \citep{haus_radiative_2010,kappel_refinements_2012,haus_self-consistent_2013}. Therefore, although the model shows sensitivity to the temperature at some altitude levels at 1.7 and 2.3 $\mathrm{\mu}$m windows, we relied on a priori VIRA profiles below 55 km and CO$_2$ continuum at these bands were considered as free parameters instead. On the other hand, Figure \ref{fig:jacobians} highlights that every aerosol mode contributes to the detected radiance below 2.3 $\mathrm{\mu}$m and that, if aerosol modes base altitudes were modified, they could all contribute to the entire spectral range. Hence, we decided to perform a third retrieval in which the entire spectrum is used to retrieve the four aerosol modes vertical distributions together with the temperature profile. In summary, the burn-in retrievals can be described in the following steps:
\begin{itemize}
    \item Step 1: Mesospheric temperature (55-80 km) and modes 1 and 2 vertical distribution using 3.5-5.2 $\mathrm{\mu}$m.
    \item Step 2: Modes 2' and 3 vertical distribution and CO$_2$ continuum using 1.74 and 2.3 $\mathrm{\mu}$m windows. 
    \item Step 3: Mesospheric temperature (55-80 km), four aerosol modes vertical distribution and CO$_2$ continuum using the entire spectral range. A priori values are taken from steps 1 and 2.
\end{itemize}

In these burn-in retrievals, the five parameters defining the vertical distribution of each aerosol mode were included as free parameters and their retrieved values were also used as a priori for MultiNest retrievals. With these, we aimed at studying whether we can obtain meaningful information from every aerosol mode and whether all of these parameters are indeed necessary to be included in the retrieval process. We studied first from which of the four aerosol modes we can obtain information. This was done by running MultiNest retrievals to obtain the Bayesian evidence of different models in which different combinations of aerosol modes vertical distributions were considered. We will refer to each of these combinations (e.g., [2,2'], [2,2',3], etc.) as a `set of aerosol modes'. All of the set of aerosol modes tested are shown in the header of Table \ref{table:results_set_modes}. In these retrievals, as in the burn-in ones, the five parameters defining the vertical distribution of each aerosol mode ($N_0$,$z_b$,$z_c$,$H_{up}$,$H_{lo}$) were allowed to vary.

Once the Bayesian evidences are compared following the Jeffrey's scale (Table \ref{table:jeffreys}), we can choose the most significant set of aerosol modes and proceed to determine which of these five parameters describing each mode do actually provide information on the aerosol vertical distribution. To do this, only the set of aerosol modes that obtained the highest Bayesian evidence was considered, and the rest of aerosol modes remained fixed. In this case, we started retrieving only the parameter $N_0$ of each aerosol mode and we consecutively added one additional free parameter in each retrieval. We will refer to each of these combinations (e.g., [$N_0$,$z_b$], [$N_0$,$z_b$,$z_c$], etc.) as `set of parameters'. The parameters that were not included in each set of parameters were fixed at their a priori value shown in Table \ref{table:haus} (e.g., in the set of parameters [$N_0$,$z_b$,$z_c$], $H_{up}$ and $H_{lo}$ maintain their a priori value). All of the sets of parameters tested are shown in the header of Table \ref{table:results_set_parameters}. Once the Bayesian evidences are compared, we will then be able to define a complete set of aerosol modes and parameters, i.e., a parameterization, to maximize information on the aerosol vertical distribution available in VIRTIS-M-IR measurements.

\section{Results} \label{sec:Results}
In Table \ref{table:results_set_modes}, we show the Bayesian evidence ($\ln{Z}$) for each set of aerosol modes and all locations, six of data cube VI0025\_07 and six additional observations observed throughout the mission. As described in Section \ref{sec:rt_code}, model selection procedure is based on comparing Bayesian evidences of two given models using the Bayes factor. In this table (as well as in Tables \ref{table:results_set_parameters}, \ref{table:combined_sets}), we highlight the model that achieved the highest Bayesian evidence value, but also those which are close enough to be statistically indistinguishable (i.e., according to Jeffrey's scale, with a Bayes factor lower than 1 with respect to the best one). As an example, in the `dark' mid-latitude location of data cube VI0025\_07, the highest Bayesian evidence was achieved when retrieving aerosol modes 2, 2' and 3 (note that $\ln{Z}$ values are negative). On the other hand, when we retrieved only modes 2 and 3, we obtained a Bayesian evidence that is only 0.8 lower than the highest one. The same occurs when the four aerosol modes were retrieved, with a difference of 0.1 with respect to [2,2',3]. Therefore, the three models are highlighted. Nevertheless, in this work, our objective is to define a common parameterization for every region of the atmosphere to retrieve the aerosol vertical distribution, hence, we must choose the set that obtained the highest evidence with as few free parameters as possible in most cases. The set of aerosol modes that is included in the leading group for most cases is  [2,2',3]. For this reason, we decided to choose it as the set of aerosol modes from which it is possible to obtain information from VIRTIS measurements. Also, we want to note that including mode 1 in the retrieval process did not significantly increase the Bayesian evidence in any case.

To determine the set of aerosol modes with the highest evidence for each location, we retrieved all of the parameters that define the vertical distribution of each of them (i.e., [$N_0$, $z_b$, $z_c$, $H_{up}$, $H_{lo}$]), however, we wanted to investigate whether all of them have an actual impact on the Bayesian evidence. To do so, we ran several MultiNest retrievals in which different sets of these parameters were allowed to vary, in particular for the set of aerosol modes [2,2',3]. From the results shown in Table \ref{table:results_set_parameters}, both $N_0$ and $z_b$ must be included in the set of parameters, as they are included in all the leading combinations but one. However, the set of parameters [$N_0$,$z_b$,$z_c$] shows higher evidence than [$N_0$,$z_b$] in three cases and it is in fact the set that obtained the highest Bayesian evidence in most cases with a fewer number of free parameters. Therefore, in order not to exclude these cases and to define a single set of parameters to retrieve the aerosols vertical distribution, we propose the set of parameters [$N_0$,$z_b$,$z_c$], although we acknowledge that, in this case, the decision is not as clear as in the preceding case.

In Table \ref{table:combined_sets}, we show a comparison between the Bayesian evidence values obtained using the complete set of parameters, the reduced version (in which $H_{up}$ and $H_{lo}$ were excluded from the retrieval process) and `mode factors' (MF$_j$ where $j=1,2,2',3$) \citep{haus_self-consistent_2013}. These factors are altitude-independent multiplicative values that affect the number density of the particles but not the altitude distribution, which have been widely used to retrieve aerosol properties \citep{grassi_venus_2014,haus_self-consistent_2013,haus_atmospheric_2014,magurno_retrieval_2017,satoh_venus_2021}. We used, as a priori probability distribution function of these parameters, Gaussian functions centered in MF$_j$=$1.0$ with standard deviations of 0.5. Note that retrieving `mode factors' does not differ from retrieving just the $N_0$ parameter (i.e., the set of parameters [$N_0$]) since they both act as a multiplicative factor for the entire vertical distribution (see Eq. \ref{eq:haus}). However, we wanted to compare the Bayesian evidence obtained using the parameterization proposed here with a simpler one, in which all the aerosol modes are considered and modified by just one parameter. In every case, the Bayes factor between the complete set of parameters and the reduced version is lower than 1, showing that there is no statistical support to include $H_{up}$ and $H_{lo}$ as free parameters. Furthermore, there is strong evidence for retrieving $[N_0,z_b,z_c]$ instead of `mode factors' in all locations.

\begin{figure}
    \centering
    \includegraphics[width=1.0\linewidth]{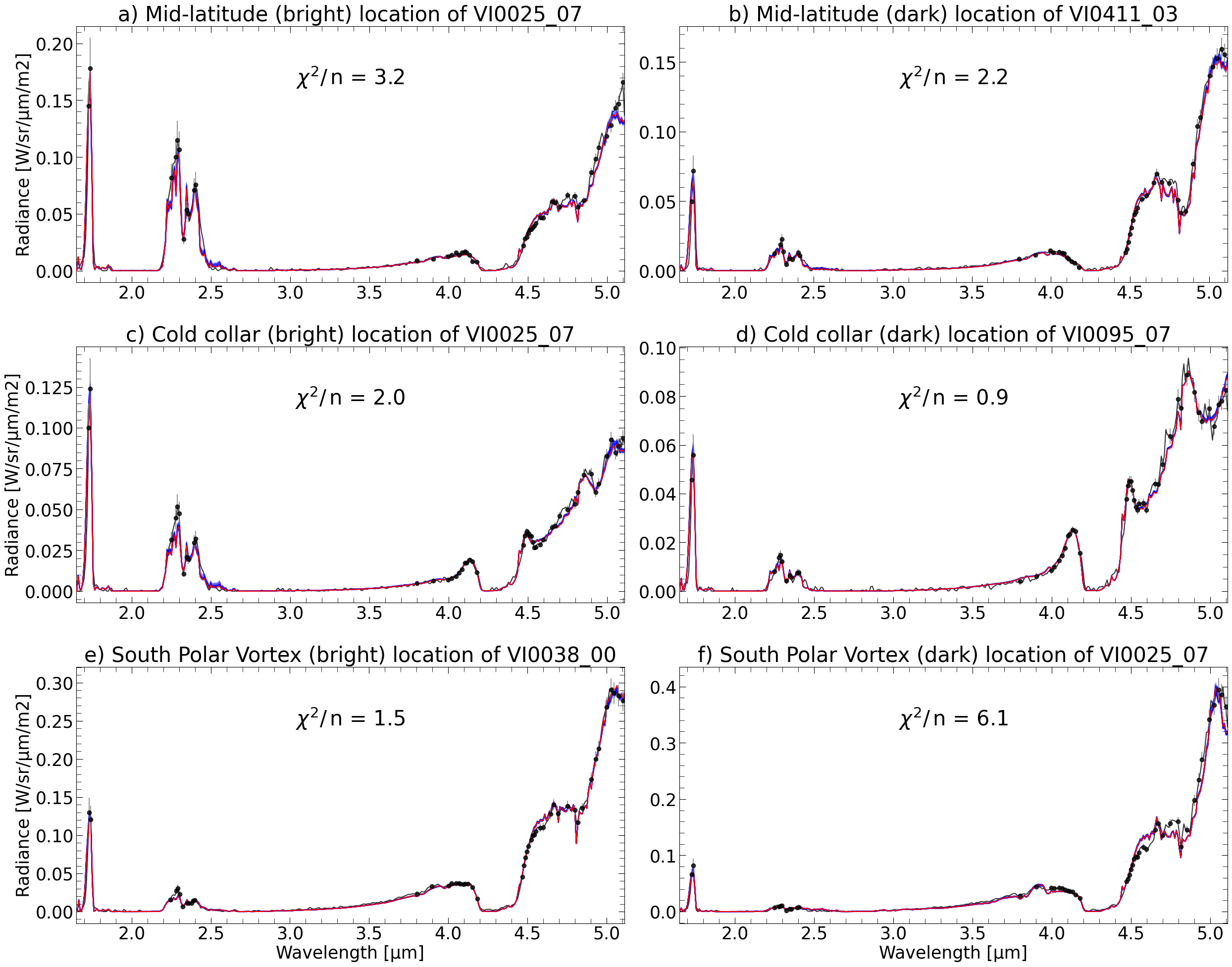}
    \caption{Best-fit (red) and observed (black) spectra of six of the studied locations including modes 2, 2' and 3 in the retrieval process with $N_0$, $z_b$ and $z_c$ as free parameters. The spectra of the best 1\% of evaluated models are represented in blue. The $\chi^2/ \mathrm{n}$ value achieved by the best-fit is shown in each panel.}
    \label{fig:spectra}
\end{figure}

As detailed in Section \ref{sec:rt_code}, we can assume that a good fit is found whenever $\chi^{2}/\mathrm{n}$ is close to 1, being n the number of spectral points. In Figure \ref{fig:spectra}, we show representative fits for mid-latitude, cold collar and South Polar Vortex regions, under `bright' and `dark' conditions, where spectra from low (e.g., panel d) to high (e.g., panel f) $\chi^{2}/\mathrm{n}$ values are displayed. While there is a range for $\chi^{2}/\mathrm{n}$ values, the average value for all cases in this work is 2.1. We want to emphasize that we are trying to fit individual spectra in this paper, which is quite challenging compared to fitting spatially binned spectra, which ensures a higher SNR and probably erases potential random errors in the measurements. Including some other free parameters might help to improve the quality of the fit, such as the CO, which has a 1-0 fundamental band in the range 4.52-4.8 $\mathrm{\mu}$m and is usually retrieved together with temperature \citep{irwin_spatial_2008,grassi_venus_2014}. In this work, however, we wanted to focus on the parameters describing the aerosols vertical distribution and model comparison using Bayesian evidence can be helpful deciding an appropriate number of free parameters when the models achieve a similar fit to the data, since, in that case, the Bayes factor is not dominated by the likelihood.

\begin{figure}
    \centering
    \includegraphics[width=1\linewidth]{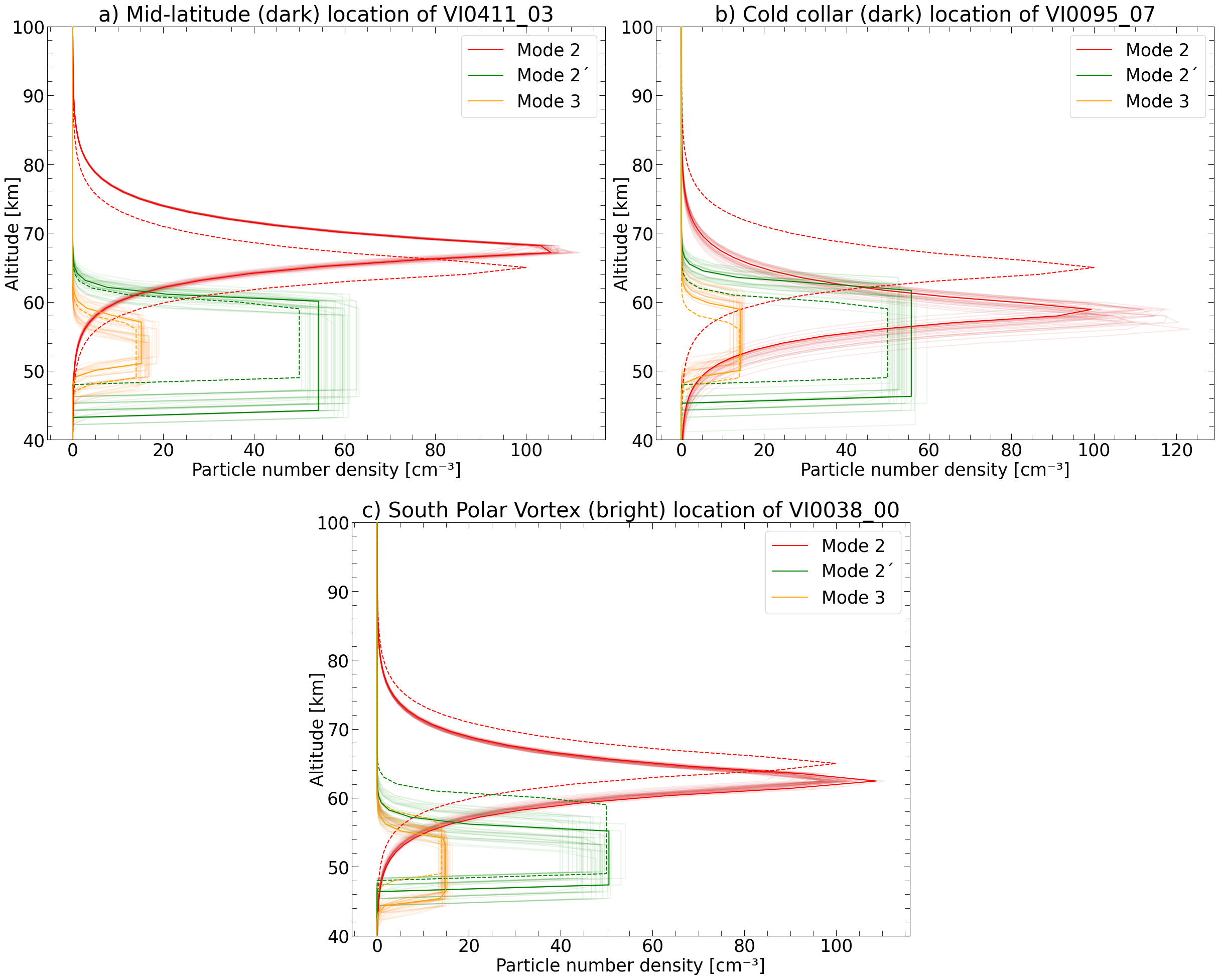}
    \caption{Best model (continuous) and \cite{haus_radiative_2016} a priori (dashed) aerosol vertical distributions of three of the studied locations including modes 2, 2' and 3 in the retrieval process with $N_0$, $z_b$ and $z_c$ as free parameters. The aerosol vertical distribution of the best 1\% of evaluated models are represented in low opacity lines.}
    \label{fig:aerosols}
\end{figure}

Figure \ref{fig:aerosols} shows the vertical aerosol distributions retrieved for same locations of panels a,c,e of Figure \ref{fig:spectra}. The vertical distribution of mode 2 particles presents a particle number density peak value similar to the \cite{haus_radiative_2016} prior although this peak is located at different altitudes: lower in the cold collar and South Polar Vortex locations and higher in the mid-latitude location. For all remaining locations, not shown in Figure \ref{fig:aerosols}, a similar general behavior was found. Middle and lower clouds present higher variability, but, in most locations, both mode 2' and 3 particles were retrieved at a slightly lower base altitude and, in some locations, they were retrieved with a higher layer thickness. We also show, in Figure \ref{fig:aerosols}, the vertical distribution of the best 1\% among all the evaluated models (corresponding to blue spectra in Figure \ref{fig:spectra}). Modes 2' and 3 present higher dispersions than mode 2, which can give us an idea of the robustness of the retrieval of each aerosol mode. 
In general terms, the shape of each aerosol mode vertical distribution is well maintained with respect to the prior in every location, with vertically extended modes 2' and 3 and a vertically concentrated mode 2, although, mode 2' particles were retrieved in a thicker layer in some cases, and we found noticeable variations in the base altitude of mode 2 particles.

CO$_2$ continuum values were not retrieved in MultiNest retrievals but in burn-in retrievals with the Optimal Estimator scheme included in archNEMESIS. The mean retrieved value for the 1.74 $\mathrm{\mu}$m window is 4.76 $\times 10^{-9}$ cm$^{-1}$ amagat$^{-2}$ which is in agreement with the continuum value used in other works \citep{de_bergh_water_1995,kappel_refinements_2012,haus_self-consistent_2013}. For the 2.3 $\mathrm{\mu}$m window, the retrieved mean value is 13.6 $\times 10^{-9}$ cm$^{-1}$ amagat$^{-2}$, which is somewhat lower than the values retrieved by \cite{pollack_near-infrared_1993} and \cite{kappel_refinements_2012}. These comparisons, however, must be taken just as indicative of our results, since retrieved continuum values depend on the utilized line database, shape and cutoff distance, and these average values are obtained from just 12 locations.

\begin{table}
\resizebox{\textwidth}{!}{%
\begin{tabular}{SlScScScScSc}
\toprule
DATA CUBE VI0025\_07       & [2,3]     & [2,2']    & [2',3]   & [2,2',3]                                       & [1,2,2',3]  \\ \midrule
Mid-latitude (bright)       & \textbf{-83.9 $\pm$ 0.1} & -86.0 $\pm$ 0.1 & -90.8 $\pm$ 0.1 & -85.1 $\pm$ 0.1 & -85.1 $\pm$ 0.1                                                                     \\ 
Mid-latitude (dark)         & \textbf{-77.2 $\pm$ 0.1} & -77.6 $\pm$ 0.1 & -80.8 $\pm$ 0.1 & \textbf{-76.4 $\pm$ 0.1} & \textbf{-76.5 $\pm$ 0.1}                                                                           \\ 
Cold collar (bright)       & -55.9 $\pm$ 0.1 & -53.7 $\pm$ 0.1 & -156.5 $\pm$ 0.1 & \textbf{-50.6$\pm$ 0.1} & \textbf{-50.8 $\pm$ 0.1}                                                                   \\ 
Cold collar (dark)         & -29.9 $\pm$ 0.1 & -29.8 $\pm$ 0.1 & -385.5 $\pm$ 0.2 & \textbf{-28.2 $\pm$ 0.1} & \textbf{-28.0 $\pm$ 0.1}                                                                     \\ 
South Polar Vortex (bright) & -167.9 $\pm$ 0.1 & -198.1 $\pm$ 0.1 & -161.6 $\pm$ 0.1 & \textbf{-105.5 $\pm$ 0.1}    & \textbf{-104.8 $\pm$ 0.1}  \\ 
South Polar Vortex (dark)   & -148.0 $\pm$ 0.2 & -157.2 $\pm$ 0.1 & -181.1 $\pm$ 0.2 & \textbf{-116.0 $\pm$ 0.2}  & \textbf{-115.3 $\pm$ 0.2} \\ \midrule
ADDITIONAL OBSERVATIONS    & \textbf{}        & \textbf{}        & \textbf{}        & \textbf{}                                             & \textbf{}                                                                               \\ \midrule
Mid-latitude (bright) & \textbf{-71.5 $\pm$ 0.1} & \textbf{-71.6 $\pm$ 0.1}  & \textbf{-70.6 $\pm$ 0.1}  & \textbf{-71.8 $\pm$ 0.1} & \textbf{-71.8 $\pm$ 0.1}  \\ 
Mid-latitude (dark) & -61.6 $\pm$ 0.1  & \textbf{-59.8 $\pm$ 0.1} & -97.3 $\pm$ 0.2  & \textbf{-60.8 $\pm$ 0.1} & \textbf{-60.7 $\pm$ 0.1} \\ 
Cold collar (bright)  & \textbf{-32.1 $\pm$ 0.1}   & \textbf{32.8 $\pm$ 0.1}  & -41.5 $\pm$ 0.1  & \textbf{-31.8 $\pm$ 0.1}   & \textbf{-31.9 $\pm$ 0.1} \\ 
Cold collar (dark)  &  -27.3 $\pm$ 0.1 & -27.7 $\pm$ 0.2   & 242.1 $\pm$ 0.1  & \textbf{-24.2 $\pm$ 0.1} & \textbf{-24.1  $\pm$ 0.1} \\ 
South Polar Vortex (bright) & -38.4 $\pm$ 0.1 & -40.7 $\pm$ 0.1  & -83.2 $\pm$ 0.1 & \textbf{-37.3 $\pm$ 0.1} & \textbf{-37.0 $\pm$ 0.1} \\ 
South Polar Vortex (dark)   & \textbf{-39.5 $\pm$ 0.1}  &  -47.4 $\pm$ 0.2 & -160.8 $\pm$ 0.2 & \textbf{-39.4 $\pm$ 0.1} & \textbf{-38.9 $\pm$ 0.1}  \\ \bottomrule          
\end{tabular}}
\caption{Bayesian evidences obtained for each tested model and location to determine the set of aerosol modes. Column headers indicate the aerosol modes included in the retrieval in each case. Boldface is used to highlight the model with the highest Bayesian evidence and models with a Bayes factor lower than 1 respect to it.}
\label{table:results_set_modes}
\end{table}

\begin{table}
\resizebox{\textwidth}{!}{%
\begin{tabular}{SlScScScScScScSc}
\toprule
DATA CUBE VI0025\_07        & [$N_0$]     & [$N_0$,$z_b$]        & [$N_0$,$z_b$,$z_c$]   & [$N_0$,$z_b$,$z_c$,$H_{up}$] & [$N_0$,$z_b$,$z_c$,$H_{lo}$] & [$N_0$,$z_b$,$z_c$,$H_{up}$, $H_{lo}$] \\ \midrule
Mid-latitude (bright)   & -90.6 $\pm$ 0.1 & \textbf{-85.6 $\pm$ 0.1}  & \textbf{-85.4 $\pm$ 0.1}& \textbf{-85.4 $\pm$ 0.1}  & \textbf{-85.0 $\pm$ 0.1}  & \textbf{-85.1 $\pm$ 0.1}     \\
Mid-latitude (dark)   & -80.3 $\pm$ 0.1 & \textbf{-76.0 $\pm$ 0.1}  & \textbf{-76.6 $\pm$ 0.1} & \textbf{-76.8 $\pm$ 0.1}  & \textbf{-76.4 $\pm$ 0.1}    & \textbf{-76.4 $\pm$ 0.1} \\
Cold collar (bright)   & -172.6 $\pm$ 0.3 & \textbf{-50.8 $\pm$ 0.1} & \textbf{-50.8$\pm$ 0.1} & \textbf{-51.1 $\pm$ 0.2}  & \textbf{-50.8 $\pm$ 0.2}    & \textbf{-50.6 $\pm$ 0.1} \\
Cold collar (dark)    & -114.5 $\pm$ 0.4 & \textbf{-28.7 $\pm$ 0.1} & \textbf{-28.3 $\pm$ 0.1}   & \textbf{-28.2 $\pm$ 0.1} & \textbf{-28.3 $\pm$ 0.1}   & \textbf{-28.3 $\pm$ 0.1}  \\
South Polar Vortex (bright) & -234.7 $\pm$ 0.3 & -107.8 $\pm$ 0.1  & \textbf{-105.7  $\pm$ 0.1} &\textbf{-105.3 $\pm$ 0.1 } & \textbf{-105.7 $\pm$ 0.1}  & \textbf{-105.5 $\pm$ 0.1} \\
South Polar Vortex (dark)   & -171.1 $\pm$ 0.2 & \textbf{-116.2 $\pm$ 0.2}  & \textbf{-116.7$\pm$ 0.2} & \textbf{-115.8 $\pm$ 0.2} & \textbf{-116.2 $\pm$ 0.2} & \textbf{-116.0 $\pm$ 0.1}  \\ \midrule
ADDITIONAL OBSERVATIONS     &   \textbf{} &    \textbf{}  &    \textbf{}       & \textbf{}   &  \textbf{}   &   \textbf{}    \\ \midrule
Mid-latitude (bright)       & \textbf{-69.2 $\pm$ 0.1} & -71.8 $\pm$ 0.1          & -72.0 $\pm$ 0.1   & -72.0 $\pm$ 0.1    & -71.7 $\pm$ 0.1    & -71.8 $\pm$ 0.1  \\
Mid-latitude (dark)    & -82.3 $\pm$ 0.2 & \textbf{-61.1 $\pm$ 0.1} & \textbf{-61.0$\pm$ 0.2}  & \textbf{-60.9 $\pm$ 0.1}  & \textbf{-60.8 $\pm$ 0.1}  & \textbf{-60.8 $\pm$ 0.1} \\
Cold collar (bright) & -58.6 $\pm$ 0.1 & \textbf{-31.8 $\pm$ 0.1}     & \textbf{-32.0 $\pm$ 0.1}  & \textbf{-31.9 $\pm$ 0.1}  & \textbf{-32.2 $\pm$ 0.1} & \textbf{-31.8 $\pm$ 0.1}  \\
Cold collar (dark)    & -123.2 $\pm$ 0.4 & -26.7 $\pm$ 0.1 & \textbf{-24.3 $\pm$ 0.1}& \textbf{-24.3 $\pm$ 0.1}& \textbf{-24.1 $\pm$ 0.1   }& \textbf{-24.2 $\pm$ 0.1}  \\
South Polar Vortex (bright) & -55.6 $\pm$ 0.2 & \textbf{-37.0 $\pm$ 0.1}   & \textbf{-37.4 $\pm$ 0.1}  & \textbf{-37.2 $\pm$ 0.1}   & \textbf{-37.3 $\pm$ 0.1}  & \textbf{-37.3 $\pm$ 0.1}  \\
South Polar Vortex (dark)   & -79.3 $\pm$ 0.3    & -44.7 $\pm$ 0.2   &\textbf{-39.4 $\pm$ 0.1} & \textbf{-39.5 $\pm$ 0.1}   & \textbf{-39.5 $\pm$ 0.1}  &\textbf{-39.4 $\pm$ 0.1}     \\ \bottomrule
\end{tabular}}
\caption{Bayesian evidences obtained for each tested model and location to determine the set of parameters, using the set of aerosol modes [2,2',3] in the retrieval. Column headers indicate the parameters that were allowed to vary in each case. Boldface is used to highlight the model with the highest Bayesian evidence and models with a Bayes factor lower than 1 respect to it.}
\label{table:results_set_parameters}
\end{table}

\begin{table}
\resizebox{\textwidth}{!}{%
\begin{tabular}{SlScScScScScSc}
\hline
DATA CUBE VI0025\_07          & [$N_0$,$z_b$,$z_c$,$H_{up}$, $H_{lo}$]  & [$N_0$,$z_b$,$z_c$]                                  & \multicolumn{1}{c}{[$MF_1$,$MF_2$,$MF_{2'}$,$MF_3$]} & Bayes factor & \multicolumn{1}{c}{Cloud top [km]} & Cloud Opacity \\ \hline
Mid-latitude (bright)         & \textbf{-85.1 $\pm$ 0.1}       & \textbf{-85.4 $\pm$ 0.1} & -89.8 $\pm$ 0.1                          & 4.4                   & 72 $^{+1}_{-1}$  & 31 $^{+7}_{-5}$       \\
Mid-latitude (dark)           & \textbf{-76.4 $\pm$ 0.1}      & \textbf{-76.6 $\pm$ 0.1} & -82.2 $\pm$ 0.1                          & 5.6                   &  71 $^{+1}_{-1}$  & 39 $^{+8}_{-7}$       \\
Cold collar (bright)          & \textbf{-50.6 $\pm$ 0.1}     & \textbf{-50.8 $\pm$ 0.1} & -62.6 $\pm$ 0.2                          & 11.8                   &  65 $^{+3}_{-1}$ & 36 $^{+7}_{-6}$    \\
Cold collar (dark)            & \textbf{-28.3 $\pm$ 0.1}      & \textbf{-28.3 $\pm$ 0.1} & -40.6 $\pm$ 0.2                          & 12.3                  &  65 $^{+2}_{-1}$  & 42 $^{+8}_{-7}$  \\
South Polar Vortex (bright)  & \textbf{-105.5 $\pm$ 0.1}      & \textbf{-105.7 $\pm$ 0.1} & -203.0 $\pm$ 0.1                          & 97.8                     &  67 $^{+1}_{-1}$  & 32 $^{+7}_{-6}$ \\
South Polar Vortex (dark)    & \textbf{-116.0 $\pm$ 0.1}      & \textbf{-116.7 $\pm$ 0.1} & -159.7 $\pm$ 0.1                        & 43                   &  68 $^{+1}_{-1}$  & 41 $^{+8}_{-6}$ \\ \hline
ADDITIONAL OBSERVATIONS       & \textbf{}             & \textbf{}                                             &                                           &                       &                                                 \\ \hline
Mid-latitude (bright)         & \textbf{-71.8 $\pm$ 0.1}      & \textbf{-72.0 $\pm$ 0.1} &  \textbf{-71.3 $\pm$ 0.1}                                    & -0.7     &  71 $^{+1}_{-1}$  &  31 $^{+7}_{-6}$        \\
Mid-latitude (dark)          & \textbf{-60.8 $\pm$ 0.1}      & \textbf{-61.0 $\pm$ 0.1} &   -69.5 $\pm$ 0.1                       & 8.5  &  73 $^{+1}_{-1}$   & 39 $^{+8}_{-7}$       \\
Cold collar (bright)          & \textbf{-31.8 $\pm$ 0.1}      & \textbf{-32.0 $\pm$ 0.1} &  -45.6 $\pm$ 0.1                         & 13.6                   & 68 $^{+1}_{-2}$   &31 $^{+7}_{-6}$   \\
Cold collar (dark)            & \textbf{-24.2 $\pm$ 0.1}      & \textbf{-24.3 $\pm$ 0.1} &  -38.0 $\pm$ 0.1                 & 13.7              &  66 $^{+1}_{-1}$   & 48 $^{+8}_{-8}$ \\
South Polar Vortex (bright)  & \textbf{-37.3 $\pm$ 0.1}      & \textbf{-37.4 $\pm$ 0.1} & -42.8 $\pm$ 0.1                     & 5.4                   &  68 $^{+1}_{-1}$   & 36 $^{+7}_{-6}$   \\
South Polar Vortex (dark)     & \textbf{-39.4 $\pm$ 0.1}      & \textbf{-39.4 $\pm$ 0.1} & -48.6 $\pm$ 0.1                  & 9.2     &  66 $^{+2}_{-1}$    & 53 $^{+9}_{-8}$                                           \\ \hline
     & \multicolumn{1}{l}{}  & \multicolumn{1}{l}{}                                  &                                           & \multicolumn{1}{l}{}  &                                                
\end{tabular}}
\caption{Bayesian evidence comparison between models using the best sets of modes [2,2',3] with the complete set of parameters, the reduced version without $H_{up}$ and $H_{lo}$ and `mode factors' parameterization. Bayes factor values compares the model with the reduced set of parameters and `mode factors' parameterization. Boldface is used to highlight the model with the highest Bayesian evidence and models with a Bayes factor lower than 1 respect to it. Cloud top and total cloud opacity values correspond to the reduced set of parameters at 1 $\mathrm{\mu} $m.} 
\label{table:combined_sets}
\end{table}

\begin{figure}
    \centering
    \includegraphics[width=1\linewidth]{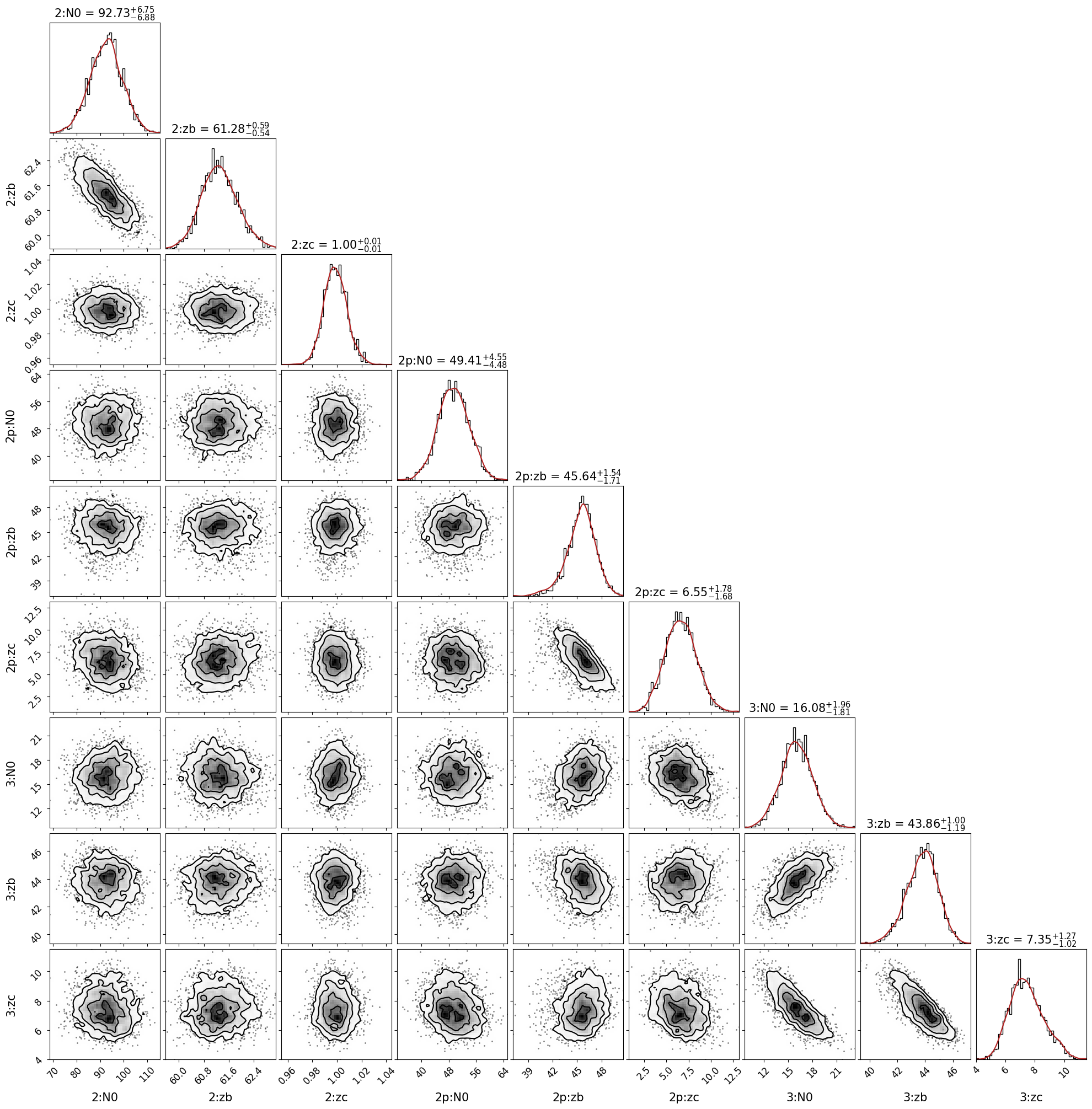}
    \caption{Corner plot of the retrieval for the `bright' South Polar Vortex location of data cube VI0025\_07. Marginalized posterior probability distribution functions of each parameter are shown in the upper panel of each column. The red lines show kernel density estimation smoothing of the functions with a Gaussian kernel. The rest of the plots show the marginalized posterior probability distributions as function of a pair of parameters. Units of each parameter are indicated in Table \ref{table:determination_parameters}.}
    \label{fig:corner_polarvortex}
\end{figure}

\section{Discussion} \label{sec:Discussion}

\subsection{Parameter determination}
We will focus here on the set of modes and parameters that obtained the highest evidence in most cases, i.e., [2,2',3] and [$N_0$,$z_b$,$z_c$]. In Table \ref{table:determination_parameters}, we show, as an example, a priori (obtained in the burn-in retrieval) and retrieved values of each parameter, as well as the uncertainties obtained, for the `bright' South Polar Vortex location of data cube VI0025\_07. Retrieved values for every location are shown in Tables \ref{table:retrieval_0025_07} and \ref{table:retrieval_additional}. The highest uncertainties are found for the parameter $N_0$, with a mean relative uncertainty of 8\%. Base altitudes, $z_b$, and layer thicknesses, $z_c$, were retrieved mainly with an uncertainty of 1-2 km, significantly improving the a priori uncertainty. The retrieved values shown in these tables for each parameter are the median value of the posterior probability distribution function of each of them; analyzing them in a corner plot gives us visual information about how constrained those values are, as well as about the correlations between them and degeneracies, which will be discussed later. Since with MultiNest retrievals we do not obtain a single value but a posterior probability distribution function, retrieved uncertainty values shown in these tables are the standard deviation of the probability distribution functions ($1-\sigma$). On the other hand, although a priori values for MultiNest retrievals were obtained in the burn-in retrievals and are thus different for every location, a priori uncertainties (i.e., the standard deviations of the a priori probability distribution functions) were the same for every location. These uncertainties were calculated ensuring that the variability obtained for each parameter in all the burn-in retrievals was taken into account.
\begin{table}
\begin{tabular}{SlSlSc}
\toprule
                    & All locations                                           & South Polar Vortex (bright) \\ \midrule
                    & A priori                                              & Retrieved                            \\ \midrule 
$2$:$N_0$ [cm$^{-3}$]   & 90 $\pm$ 10                                          & 96 $^{+8}_{-7}$                      \\
$2$:$z_b$ [km]  & 62 $\pm$ 5                                            & 62.1 $\pm$ 0.5\\ 
$2$:$z_c$ [km] & 1.00 $\pm$ 0.01                                       & 1.00 $\pm$ 0.01\\ 
$2'$:$N_0$ [cm$^{-3}$]  & 50 $\pm$ 5                                            & 49 $\pm$ 4\\ 
$2'$:$z_b$ [km]& 46 $\pm$ 5                                            & 47$\pm$ 2\\ 
$2'$:$z_c$ [km]& 7 $\pm$ 3                                            & 9 $\pm$ 2\\ 
$3$:$N_0$ [cm$^{-3}$]   & 14 $\pm$ 3                                        & 16 $\pm$ 2\\ 
$3$:$z_b$ [km]& 44 $\pm$ 5                                            & 46 $^{+1}_{-2}$                \\ 
$3$:$z_c$ [km]& 8 $\pm$ 2.5                                             & 8 $\pm$ 1\\ \bottomrule
\end{tabular}
\caption{A priori value and posterior probability distribution median value of each parameter for the `bright' South Polar Vortex location of data cube VI0025\_07. }
\label{table:determination_parameters}
\end{table}

In Figure \ref{fig:corner_polarvortex}, we show a corner plot of these parameters for the `bright' South Polar Vortex location of data cube VI0025\_07. The upper panels of each column plot the marginalized posterior probability distribution function of each parameter. A Gaussian-shaped function implies that the most probable value of the parameter is well determined, with an uncertainty that can be approximately represented by the Full Width at Half Maximum (FWHM) of the function. On the other hand, a uniformly shaped function for a parameter tells us that it has multiple values that can reproduce the observations. In this particular location, all of the parameters have a Gaussian shaped probability distribution function. The corner plots for every location are included in Appendix \ref{sec:appendix_corner} and show that most of the parameters posterior probability distribution functions are Gaussian-shaped.  

It must be noted that the 1\% best models in Figure \ref{fig:spectra}, whose aerosol vertical distribution is also included in Figure \ref{fig:aerosols}, provide a fit of the data which is fairly similar, despite differences in the retrieved parameter values. This is precisely the strength of the Bayesian inference technique used here, which maps the whole posterior distribution of parameters and prevents picking a given combination of parameters without taking into account that there are many others that can reproduce the data as well.

There are other parameters that could be introduced into the retrieval process, such as aerosol particle sizes or H$_2$SO$_4$ concentration, that can present spatial variability. \cite{wilson_evidence_2008} analyzed the relative brightness of 1.74 and 2.3 $\mathrm{\mu}$m bands and suggested anomalous particles within the center of polar vortices, later confirmed by \cite{barstow_2012}, who also found higher sulfuric acid concentrations in regions of optically thick clouds. Moreover, \cite{magurno_retrieval_2017} obtained better fits with higher sulfuric acid concentrations than the widely used value of 75\% (\citeauthor{pollack_near-infrared_1993}, \citeyear{pollack_near-infrared_1993}; \citeauthor{zasova_structure_2006}, \citeyear{zasova_structure_2006}; \citeauthor{grassi_retrieval_2008}, \citeyear{grassi_retrieval_2008}, \citeyear{grassi_venus_2014}; \citeauthor{tsang_correlated-k_2008}, \citeyear{tsang_correlated-k_2008}; \citeauthor{haus_atmospheric_2014}, \citeyear{haus_atmospheric_2014}). 
In this work, we focus on the information that can be retrieved about the aerosols vertical distribution, but other parameters could be included in the future following the same scheme for parameter selection.

\subsection{Parameter correlations}
Corner plots also show the posterior probability distribution function by pairs of parameters, allowing to detect possible correlations between them. In Figure \ref{fig:corner_polarvortex}, we can see a clear anti-correlation between parameters $3$:$N_0$ and $3$:$z_c$, and between $3$:$z_b$ and $3$:$z_c$, as well as a positive correlation between $3$:$N_0$ and $3$:$z_b$. In fact, this is not an isolated case, rather, we have detected this behavior in almost every location. Anti-correlations between $2$:$N_0$ and $2$:$z_b$, and between $2'$:$z_b$ and $2'$:$z_c$, are also quite common among the studied locations. Correlations (and anti-correlations) show the dependence between the values of two parameters to achieve the same atmospheric status. For example, the anti-correlations between $N_0$ and $z_c$ in Figure \ref{fig:corner_polarvortex} show that the observations can be equally fitted by a physically thinner, denser cloud and by a more extended, less dense cloud. The sampling of the free parameter space by nested sampling reveals such correlations as well as the most probable combination of parameter values. Finally, we did not find clear correlations between parameters of different aerosol modes, which means that these data are able to properly disentangle the contribution from each aerosol mode.

\subsection{Cloud vertical structure}
It is not straightforward to compare our results with previous works, as \cite{haus_self-consistent_2013} parameterization has only been used in combination with `mode factors' discussed in previous sections. Aerosol exponential profiles in which aerosol scale height and cloud top are fitted as free parameters are also widely used. \cite{lee_vertical_2012} used this parameterization to investigate cloud top structure using VIRTIS and VEx/VeRa data and, assuming only mode 2 particles above 60 km, retrieved clouds at lower altitude levels in the cold collar region to correctly reproduce the inversion in the wings of the 4.3 $\mathrm{\mu}$m band, which is consistent with our results.

However, we can use derived magnitudes as a cross check for our results. Table \ref{table:combined_sets} shows the cloud top obtained for each location using the best set of aerosol modes [2,2',3] and the reduced version of the parameter set [$N_0$,$z_b$,$z_c$]. This magnitude is defined as the altitude at which the cumulative cloud optical depth from the top of the atmosphere reaches unity, and it is wavelength dependent. In this paper, we calculated cloud top at 1 $\mathrm{\mu}$m, for easy comparison with previous works. The retrieved cloud top values in the mid-latitude
locations analyzed here were always higher than the cold collar ones, showing a poleward decrease in cloud top altitude as most studies agree on (\citeauthor{ignatiev_altimetry_2009}, \citeyear{ignatiev_altimetry_2009}; \citeauthor{grassi_retrieval_2008}, \citeyear{grassi_retrieval_2008}; \citeauthor{lee_vertical_2012}, \citeyear{lee_vertical_2012};   \citeauthor{haus_atmospheric_2014},\citeyear{haus_atmospheric_2014}; \citeauthor{garate-lopez_instantaneous_2015}, \citeyear{garate-lopez_instantaneous_2015}). However, in the South Polar Vortex, we did not retrieve lower cloud tops compared to those obtained in the cold collar locations. The South Polar Vortex is a permanent feature of the Venusian atmosphere that presents a high variability with timescales of 1-2 days \citep{garate_lopez_2013} and, since we only analyzed here four specific locations inside it, we cannot state whether these similar cloud top values are representative or coincidental until we complete the full analysis of the VIRTIS database we are currently conducting.

The cloud opacity value at 1 $\mathrm{\mu} $m for each location is also included in Table \ref{table:combined_sets} and higher values were always found under `dark' conditions. This is consistent with \cite{arnold_venus_2008} analysis, who suggested that low radiance values at 2.3 $\mathrm{\mu} $m window (which is the band that we used to determine `bright' and `dark' conditions) may be related to high cloud thickness. \cite{haus_atmospheric_2014} found a latitudinal dependence of cloud opacity with a slight decrease from mid-latitude to cold collar latitude ranges and a poleward increase in cloud opacity. If we compare `bright' and `dark' locations independently, a slight poleward increasing trend can be observed, with the highest cloud opacity values obtained in the South Polar Vortex locations. However, cloud opacity values can display high variability and, therefore, general latitudinal dependence cannot be inferred from the 12 specific locations analyzed here.  

In our retrievals, we did not apply any constraint to the lower cloud base (LCB) altitude. This altitude can be estimated in our model as the lower altitude between $2'$:$z_b$ and $3$:$z_b$ since both modes 2' and 3 particles have a very small lower scale height $H_{lo}$. The nephelometer experiments conducted onboard the four probes of Pioneer Venus Multiprobe obtained four LCB altitudes at 47.5, 47.1, 48.5 and 50.4 km \citep{ragent_structure_1980} and radio occultation experiments onboard Pioneer Venus Orbiter \citep{cimino_composition_1982} showed that this LCB altitude ranges from 47-48 km at mid-latitudes to 43-47 km at high and subpolar latitudes. \cite{barstow_2012} retrieved, among other parameters, the latitudinal variability of the base altitude in their analysis and obtained lower LCB altitudes at subpolar latitudes. The mean LCB altitude obtained in this work is 47.3 km, which is consistent with these previous works, although enhanced LCB altitude variability at high latitudes cannot be inferred from the individuals locations analyzed here.

\begin{figure}
\centering
\includegraphics[width=.3\linewidth]{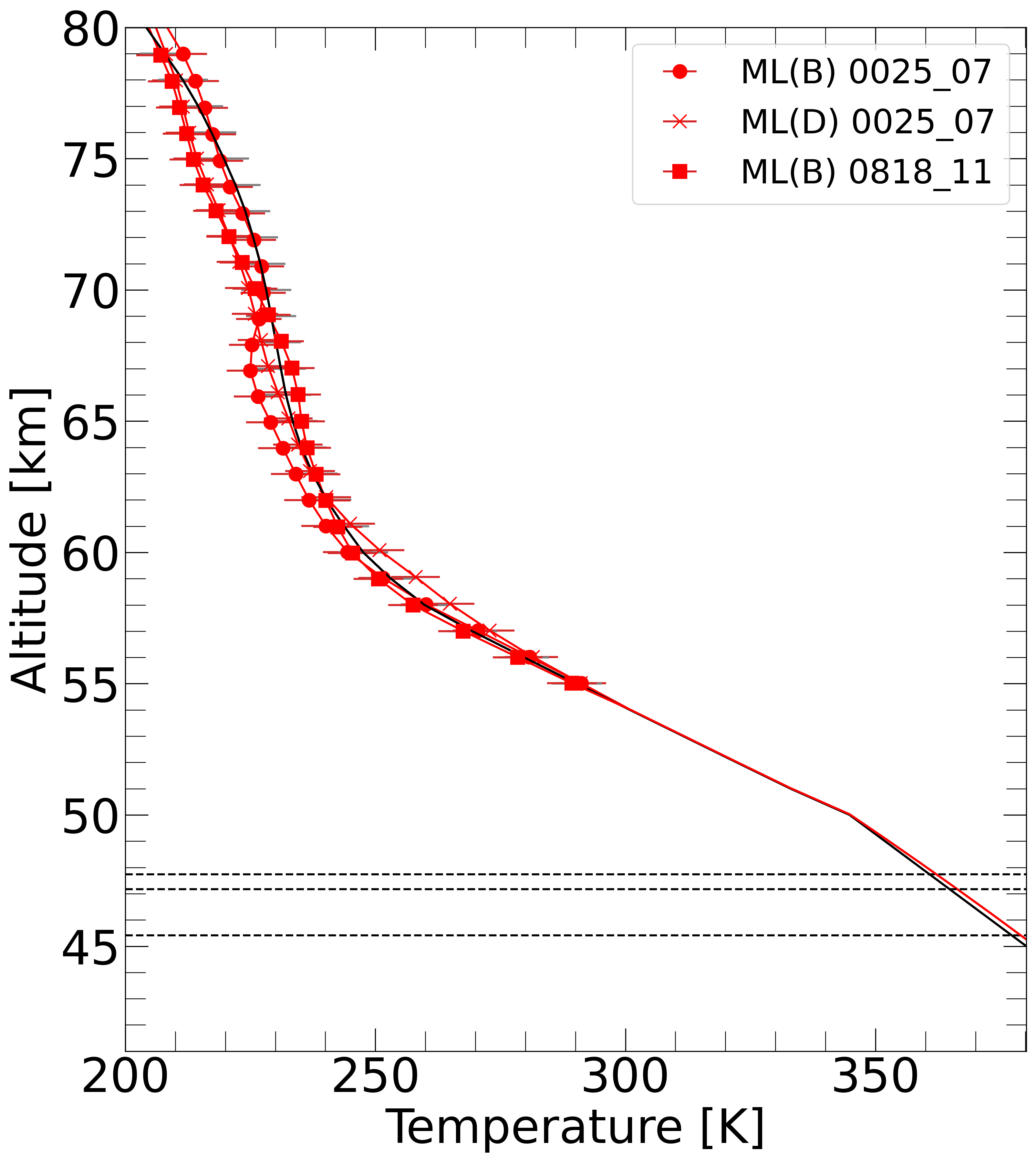}
\hfill
\centering
\includegraphics[width=.3\linewidth]{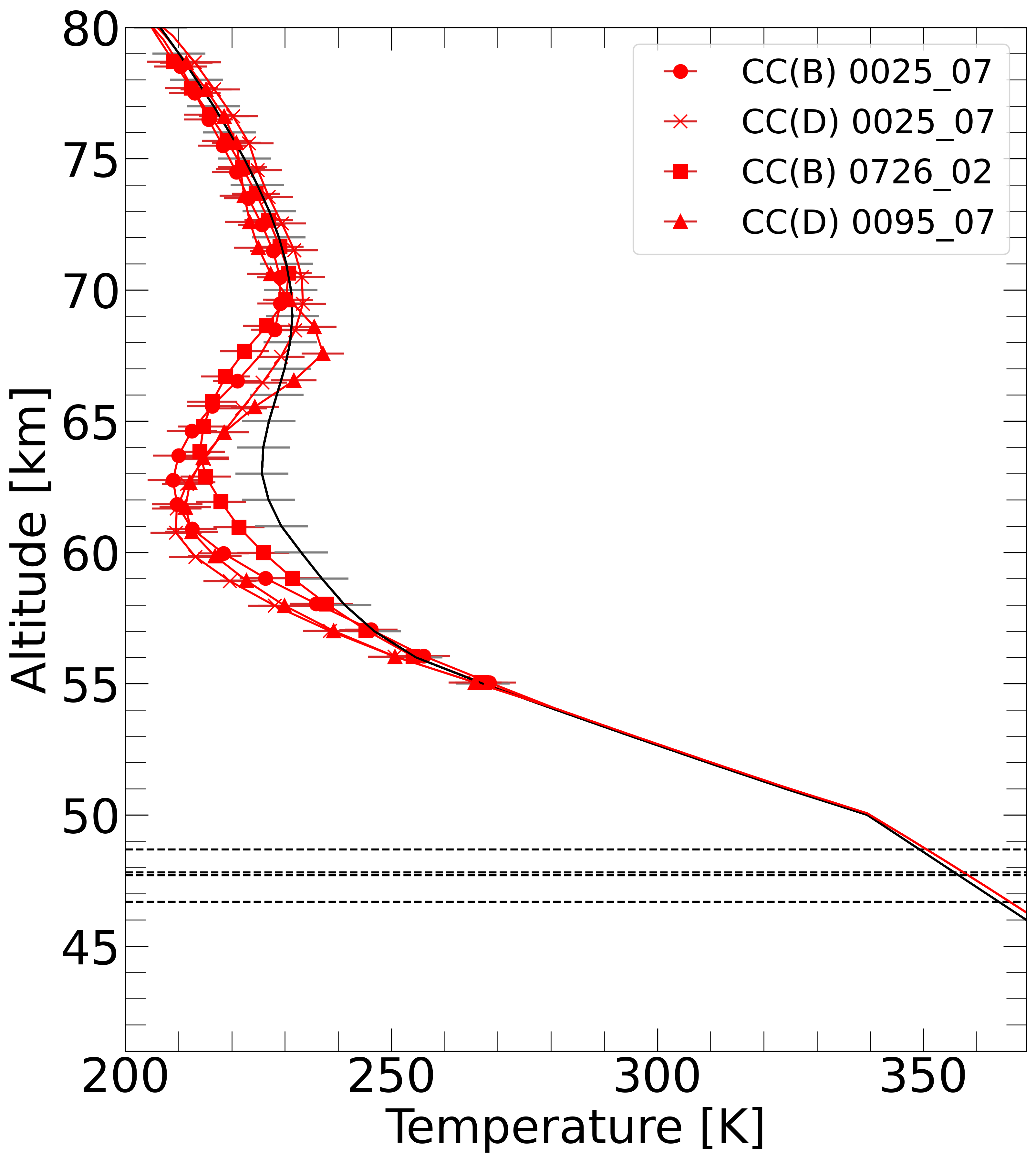}
\hfill
\centering
\includegraphics[width=.3\linewidth]{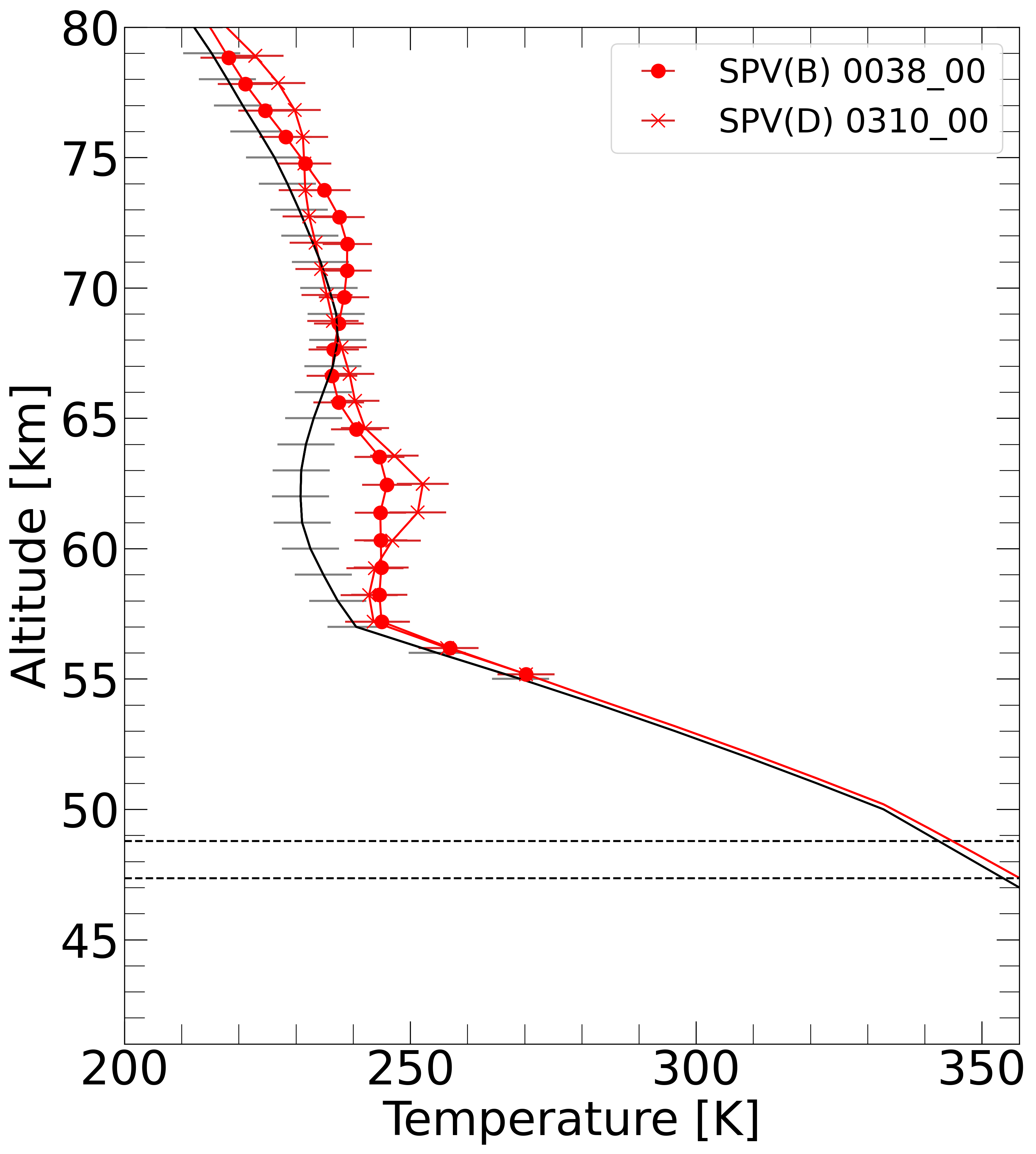}
\caption{Temperature profiles obtained with archNEMESIS optimal estimator scheme at different locations. Retrieved profiles are represented in red together and a priori profiles from VIRA dataset in black. Horizontal bars in temperature profiles show a priori and retrieved uncertainty at each altitude level. Horizontal dashed black lines show retrieved lower cloud base altitude for each case. ML, CC and SPV stand for mid-latitude, cold collar and South Polar Vortex regions, respectively. B and D stand for `bright' and `dark' conditions, respectively.}
\label{fig:temp_profiles}
\end{figure}

\subsection{Temperature vertical structure}
As we have already described in Section \ref{sec:retrievals}, temperature vertical profiles were not retrieved using MultiNest due to the large computation times required for a sensible vertical sampling of the profile. Instead, we used the archNEMESIS optimal estimator scheme to retrieve an individual profile for each studied location. In these retrievals, the a priori profiles are taken from VIRA-1 \citep{seiff_models_1985} and VIRA-2 \citep{zasova_structure_2006} datasets connected by a linear interpolation between 40 and 50 km. These datasets also exhibit solar longitude dependence of temperature profiles, but we have averaged them to use a single profile for each latitude range. An a priori uncertainty value of 5K was assumed at every altitude level between 55 and 80 km in the retrieval process. This value is important to interpret the retrieved uncertainties, since, within an optimal estimator scheme, posterior uncertainties are affected by both measurement uncertainty and a priori parameter uncertainties (see Eq. 3.31 of \citealp{rodgers_2000})

Figure \ref{fig:temp_profiles} shows temperature vertical profiles retrieved at some of the studied locations. Even though these are profiles from specific locations, general features of the average thermal structure of Venus can be observed. Thermal inversions are expected in the cold collar region at about 62 km altitude level, with temperatures values more than 15 K colder than the pole (\citeauthor{taylor_structure_1980}, \citeyear{taylor_structure_1980}; \citeauthor{zasova_structure_2006}, \citeyear{zasova_structure_2006}; \citeauthor{garate-lopez_instantaneous_2015}, \citeyear{garate-lopez_instantaneous_2015}), which is correctly reproduced by our retrieval process. A difference of 10-20 K above 70 km was obtained between the mid-latitude and South Polar Vortex profiles, which is also in agreement with the average thermal structure obtained by \cite{haus_atmospheric_2014}. The mean retrieved error in temperature is around 4 K, which is consistent with the uncertainty obtained in other works that studied the mesospheric temperature structure \citep{irwin_nemesis_2008,grassi_venus_2014,garate-lopez_instantaneous_2015}.

Cloud structure can also be tied to temperature vertical profiles by cloud microphysics since, at altitudes below $\sim$40 km, the temperature usually exceeds 400 K and the thermochemistry is governed by the destruction of sulfuric acid \citep{krasnopolsky_chemical_2007,mcgouldrick_one-dimensional_2026}. In our retrieved aerosol vertical structures, the LCB altitude was not obtained at altitudes were the retrieved temperature was higher than 380 K in any location, as can be seen in the examples shown in Figure \ref{fig:temp_profiles}, where the LCB altitude is represented with horizontal dashed black lines.

\section{Conclusions} \label{sec:Conclusions}

In this work, we have analyzed the information on the aerosol vertical distribution that can be retrieved from VIRTIS-M infrared data. Taking into account the high latitudinal variability of cloud and temperature structures, twelve specific locations have been selected to study their differences  in aerosol properties through an intensive analysis of the prior and posterior spaces of free parameters. Modes 2, 2' and 3 particles contribute substantially to the observed signal in every location, while smaller mode 1 particles appear to be less relevant. Moreover, based on Bayesian evidence results, we propose here a parameterization based on \cite{haus_self-consistent_2013} description of aerosol vertical distribution, in which the parameters $[N_0, z_b, z_c]$ are retrieved for every aerosol mode to obtain more information on the aerosol vertical distribution, instead of using scaling multiplicative factors. Bayesian evidence supports this set of free parameters as the most informative option with the fewest parameters, maximizing the return from numerical simulations. Using this modified version, the retrieved temperature and cloud structures are in general agreement with previous studies, although higher cloud top values were found inside the South Polar Vortex. We summarize the main conclusions from this study in the following:
\begin{itemize}
    \item MultiNest has been used for the first time to study aerosol properties in the atmosphere of Venus and it shows that Bayesian evidence is a powerful discriminant factor to choose among competing aerosol vertical distribution models that can reproduce the same observed spectrum.
    \item Peak particle number density, base altitude and layer thickness information of modes 2, 2' and 3 can be reliably retrieved from VIRTIS-M infrared data using \cite{haus_self-consistent_2013} aerosol vertical distribution description if individual parameters are retrieved instead of `mode factors'.
    \item Cloud top and cloud opacity values obtained are in general agreement with previous latitudinal variability studies, although higher cloud top values were found inside the South Polar Vortex. 
\end{itemize}

As we analyzed here a few individual, albeit representative, locations, our parameter results should be taken only as tentative and not necessarily extensible to other regions and times. Further analysis of the entire VIRTIS-M infrared data base is mandatory to draw general conclusions. This will be addressed in a subsequent paper in which instantaneous horizontal cloud and temperature structures will be retrieved, allowing to study their temporal variability in short timescales.

\section*{CRediT authorship contribution statement}
\noindent\textbf{Jaime Reyes-Guerrero:} Conceptualization, Data curation, Formal analysis, Investigation, Methodology, Software, Visualization, Writing - original draft, Writing - review \& editing.\\
\noindent\textbf{Santiago Pérez-Hoyos:} Conceptualization, Methodology, Formal analysis, Supervision, Writing - review \& editing.\\
\noindent\textbf{Itziar Garate-Lopez:} Conceptualization, Methodology, Formal analysis, Supervision, Writing - review \& editing.\\

\section*{Acknowledgements}
This work was supported by the Basque Government (Grupos de Investigación, IT1742-22), Elkartek KK-2025/00106 and by Grant PID2023/149055NB/C31 funded by MICIU/AEI/10.13039/501100011033 and by FEDER, UE. J. Reyes-Guerrero acknowledges a PhD scholarship from EHU. We thank Dr. Robert Hargreaves for his assistance in compiling absorption k-tables from HITEMP database. The authors acknowledge the effort of Prof. Patrick Irwin, Dr. Juan Alday, Joseph Penn, and the rest of the NEMESIS and archNEMESIS team in developing the tool, maintaining it, and providing personal support to the users. The authors acknowledge the computational resources and human support provided by the DIPC Supercomputing Center. The VIRTIS data used in this work are publicly available at the ESA Planetary Sciences Archive: https://archives.esac.esa.int/psa/ftp/VENUS-EXPRESS/VIRTIS/.

\appendix
\printcredits
\renewcommand\thefigure{\thesection.\arabic{figure}}   
\renewcommand\thetable{\thesection.\arabic{table}}
\setcounter{table}{0}
\setcounter{figure}{0}
\setcounter{page}{1}
\section{Parameter values} \label{sec:appendix_param_values}
\newpage
\begin{table*}
\resizebox{\textwidth}{!}{%
\begin{tabular}{SlSlSlSlSlSlSlSl}
\toprule
                                   &                 & ML (bright)             & ML (dark)               & CC (bright)             & CC (dark)               & \multicolumn{1}{c}{SPV (bright)} & SPV (dark)              \\ \midrule
                                   & A priori        & \multicolumn{6}{c}{Retrieved}      
                                   \\ \midrule 
$2$:$N_0$ [cm$^{-3}$]  & 100 $\pm$ 10    & 108 $^{+7}_{-8}$& 109 $\pm$ 8& 99 $\pm$ 8& 105 $^{+9}_{-8}$& 92 $\pm$ 7& 99 $\pm$ 6\\
$2$:$z_b$ [km]         & 65 $\pm$ 5      & 66.3 $^{+0.5}_{-0.4}$& 65.6 $^{+0.4}_{-0.3}$& 58 $^{+1}_{-5}$& 58 $\pm$ 1& 61.3 $^{+0.6}_{-0.5}$& 62.4 $^{+0.4}_{-0.3}$\\
$2$:$z_c$ [km]         & 1.00 $\pm$ 0.01& 1.00 $\pm$ 0.01& 1.00 $\pm$ 0.01& 1.00 $\pm$ 0.01& 1.00 $\pm$ 0.01& 1.00 $\pm$ 0.01& 1.00 $\pm$ 0.01\\
$2'$:$N_0$ [cm$^{-3}$] & 50 $\pm$ 5      & 46 $^{+5}_{-4}$& 53 $\pm$ 4& 49 $\pm$ 4& 56 $\pm$ 4& 49 $^{+5}_{-4}$& 54 $\pm$ 4\\
$2'$:$z_b$ [km]       & 49 $\pm$ 5      & 50 $\pm$ 2& 45 $\pm$ 2& 50 $\pm$ 2& 46$^{+2}_{-3}$& 46 $\pm$ 2& 42 $^{+2}_{-1}$\\
$2'$:$z_c$ [km]       & 11 $\pm$ 3      & 9 $\pm$ 2& 12 $\pm$ 2& 13 $\pm$ 2& 14 $\pm$ 2& 7 $\pm$ 2& 10 $\pm$ 2\\
$3$:$N_0$ [cm$^{-3}$]  & 14 $\pm$ 3      & 13 $\pm$ 2& 16 $\pm$ 2& 16 $\pm$ 2& 15 $^{+2}_{-1}$& 16 $^{+2}_{-1}$& 19 $\pm$ 2\\
$3$:$z_b$ [km]         & 49 $\pm$ 5      & 45 $\pm$ 2& 46 $\pm$ 2& 44 $^{+1}_{-2}$& 51 $\pm$ 2& 44 $\pm$ 1& 43 $\pm$ 1\\
$3$:$z_c$ [km]         & 8.0 $\pm$ 2.5   & 7 $\pm$ 1& 8 $\pm$ 1& 7 $\pm$ 1& 9 $\pm$ 1& 7 $\pm$ 1& 9 $\pm$ 1\\ \bottomrule
\end{tabular}}
\caption{Posterior probability distribution median of each retrieved parameter and location of datacube VI0025\_07. \cite{haus_radiative_2016} a priori values are shown for comparison. Retrieved uncertainty values are $1-\sigma$ values of posterior probability distribution functions. ML, CC and SPV stand for mid-latitude, cold collar and South Polar Vortex, respectively.}
\label{table:retrieval_0025_07}
\end{table*}

\begin{table*}
\resizebox{\textwidth}{!}{%
\begin{tabular}{SlSlSlSlSlSlSlSl}
\toprule
                           &                 & ML (bright)             & ML (dark)               & CC (bright)             & CC (dark)               & \multicolumn{1}{c}{SPV (bright)} & SPV (dark)              \\ \midrule
                           & A priori        & \multicolumn{6}{c}{Retrieved}                                                                                                                                      \\ \midrule
$2$:$N_0$ {[}cm$^{-3}${]}  & 100 $\pm$ 10    & 103 $^{+8}_{-9}$& 107 $^{+7}_{-8}$& 102 $\pm$ 8& 106 $\pm$ 8& 96 $^{+8}_{-7}$& 100 $^{+8}_{-9}$\\
$2$:$z_b$ {[}km{]}         & 65 $\pm$ 5      & 64.6 $\pm$ 0.4& 67.0 $\pm$ 0.3& 61 $\pm$ 1& 59 $\pm$ 1& 62.1 $\pm$ 0.5& 58 $^{+2}_{-3}$\\
$2$:$z_c$ {[}km{]}         & 1.00 $\pm$ 0.01 & 1.00 $\pm$ 0.01& 1.00 $\pm$ 0.01& 1.00 $\pm$ 0.01& 1.00 $\pm$ 0.01& 1.00 $\pm$ 0.01& 1.00 $\pm$ 0.01\\
$2'$:$N_0$ {[}cm$^{-3}${]} & 50 $\pm$ 5      & 49 $^{+4}_{-5}$& 55 $\pm$ 4& 48 $\pm$ 4& 60 $^{+4}_{-5}$& 49 $\pm$ 4& 60 $\pm$ 4\\
$2'$:$z_b$ {[}km{]}        & 49 $\pm$ 5      & 51 $\pm$ 2& 45 $^{+2}_{-3}$& 51 $\pm$ 2& 44$^{+3}_{-4}$& 47 $\pm$ 2& 45 $^{+2}_{-3}$\\
$2'$:$z_c$ {[}km{]}        & 11 $\pm$ 3      & 11 $\pm$ 2& 13 $\pm$ 2& 10 $\pm$ 2& 16 $\pm$ 2& 9 $\pm$ 2& 15 $\pm$ 2\\
$3$:$N_0$ {[}cm$^{-3}${]}  & 14 $\pm$ 3      & 13 $\pm$ 2& 15 $\pm$ 2& 13 $\pm$ 2& 17 $^{+2}_{-1}$         & 16 $\pm$ 2& 19 $\pm$ 2\\
$3$:$z_b$ {[}km{]}         & 49 $\pm$ 5      & 47 $\pm$ 3& 49 $^{+2}_{-3}$& 46 $\pm$ 2& 51 $\pm$ 2& 46 $^{+1}_{-2}$& 52 $^{+1}_{-2}$\\
$3$:$z_c$ {[}km{]}         & 8.0 $\pm$ 2.5   & 7 $^{+2}_{-1}$& 8 $\pm$ 1& 7$^{+2}_{-1}$& 10 $\pm$ 1& 8 $\pm$ 1& 11 $\pm$ 1\\ \bottomrule
\end{tabular}}
\caption{Posterior probability distribution median of each retrieved parameter and location of additional datacubes. \cite{haus_radiative_2016} a priori values are shown for comparison. Retrieved uncertainty values are $1-\sigma$ values of posterior probability distribution functions. ML, CC and SPV stand for mid-latitude, cold collar and South Polar Vortex, respectively.}
\label{table:retrieval_additional}
\end{table*}

\section{Corner plots} \label{sec:appendix_corner}

\begin{figure*}
    \centering
    \includegraphics[width=1\linewidth]{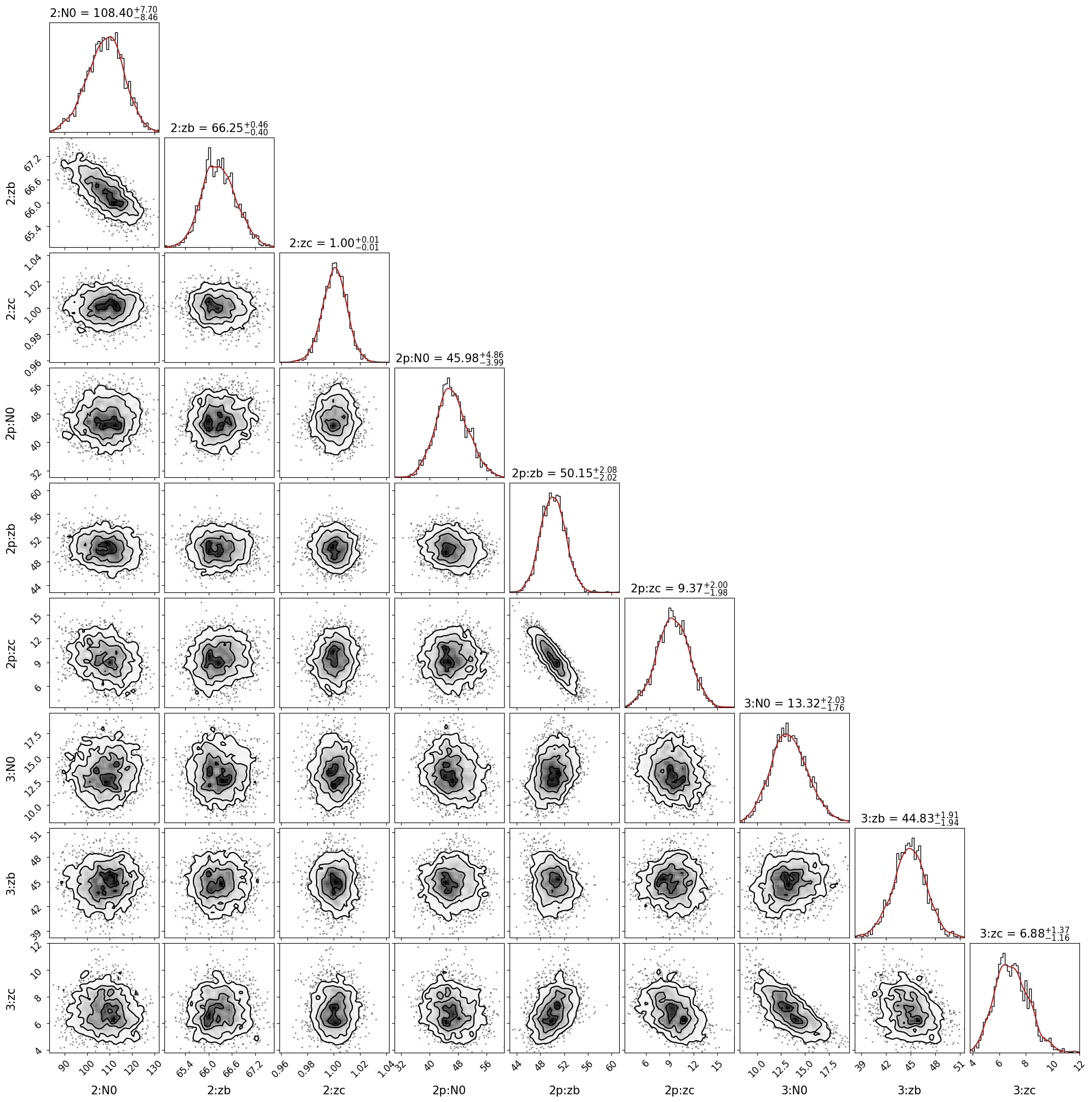}
    \caption{Corner plot of the retrieval for the `bright' mid-latitude location of data cube VI0025\_07. Marginalized posterior probability distribution functions of each parameter are shown in the upper panel of each column. The red lines show kernel density estimation smoothing of the functions with a Gaussian kernel. The rest of the plots show the marginalized posterior probability distributions as function of a pair of parameters. Units of each parameter are indicated in Table \ref{table:determination_parameters}.}
    \label{fig:corner_midlat_bright_0025_07}
\end{figure*}

\begin{figure*}
    \centering
    \includegraphics[width=1\linewidth]{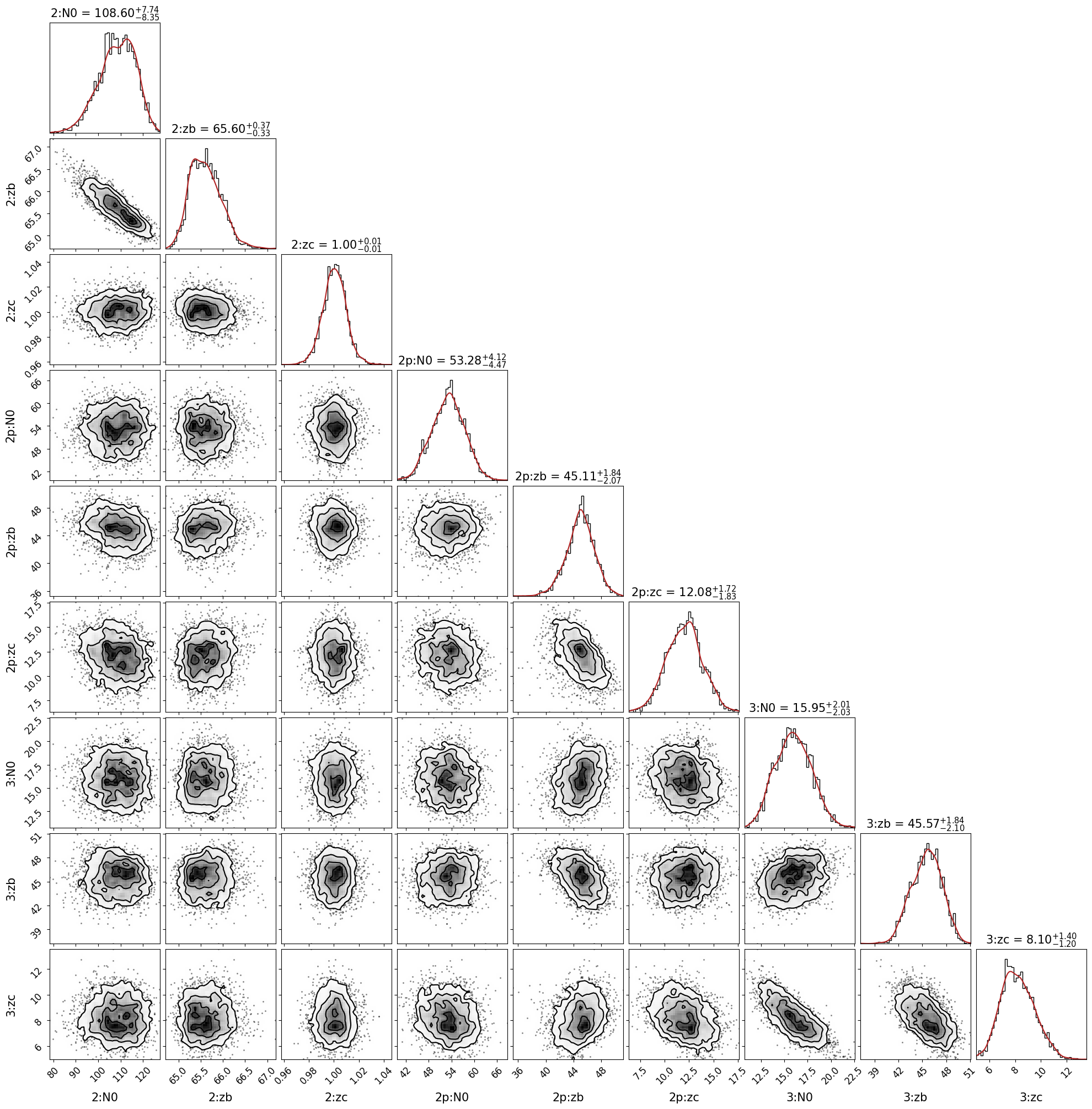}
    \caption{Corner plot of the retrieval for the `dark' mid-latitude location of data cube VI0025\_07. Marginalized posterior probability distribution functions of each parameter are shown in the upper panel of each column. The red lines show kernel density estimation smoothing of the functions with a Gaussian kernel. The rest of the plots show the marginalized posterior probability distributions as function of a pair of parameters. Units of each parameter are indicated in Table \ref{table:determination_parameters}.}
    \label{fig:corner_midlat_dark_0025_07}
\end{figure*}

\begin{figure*}
    \centering
    \includegraphics[width=1\linewidth]{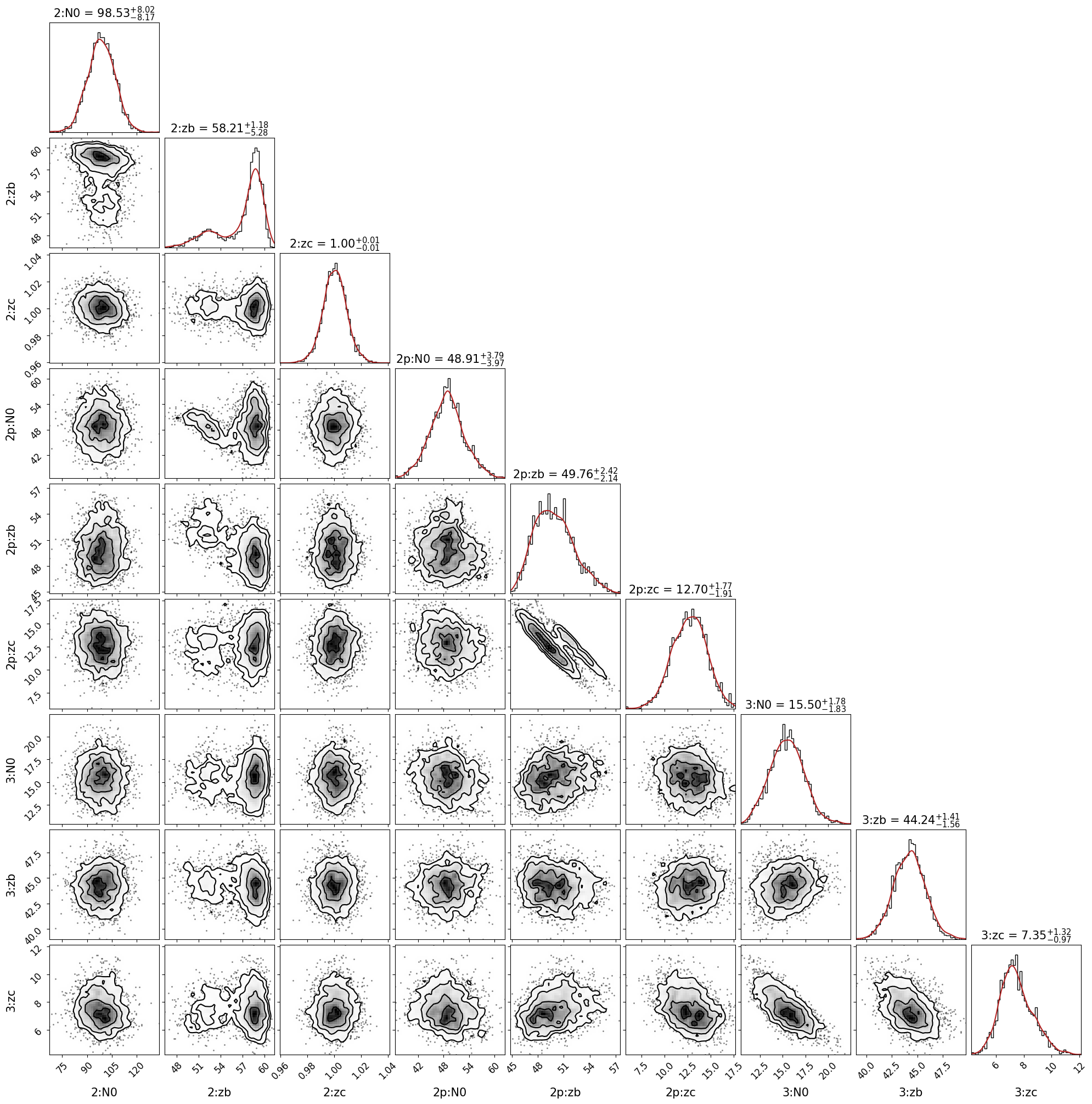}
    \caption{Corner plot of the retrieval for the `bright' cold collar location of data cube VI0025\_07. Marginalized posterior probability distribution functions of each parameter are shown in the upper panel of each column. The red lines show kernel density estimation smoothing of the functions with a Gaussian kernel. The rest of the plots show the marginalized posterior probability distributions as function of a pair of parameters. Units of each parameter are indicated in Table \ref{table:determination_parameters}.}
    \label{fig:corner_coldcollar_bright_0025_07}
\end{figure*}

\begin{figure*}
    \centering
    \includegraphics[width=1\linewidth]{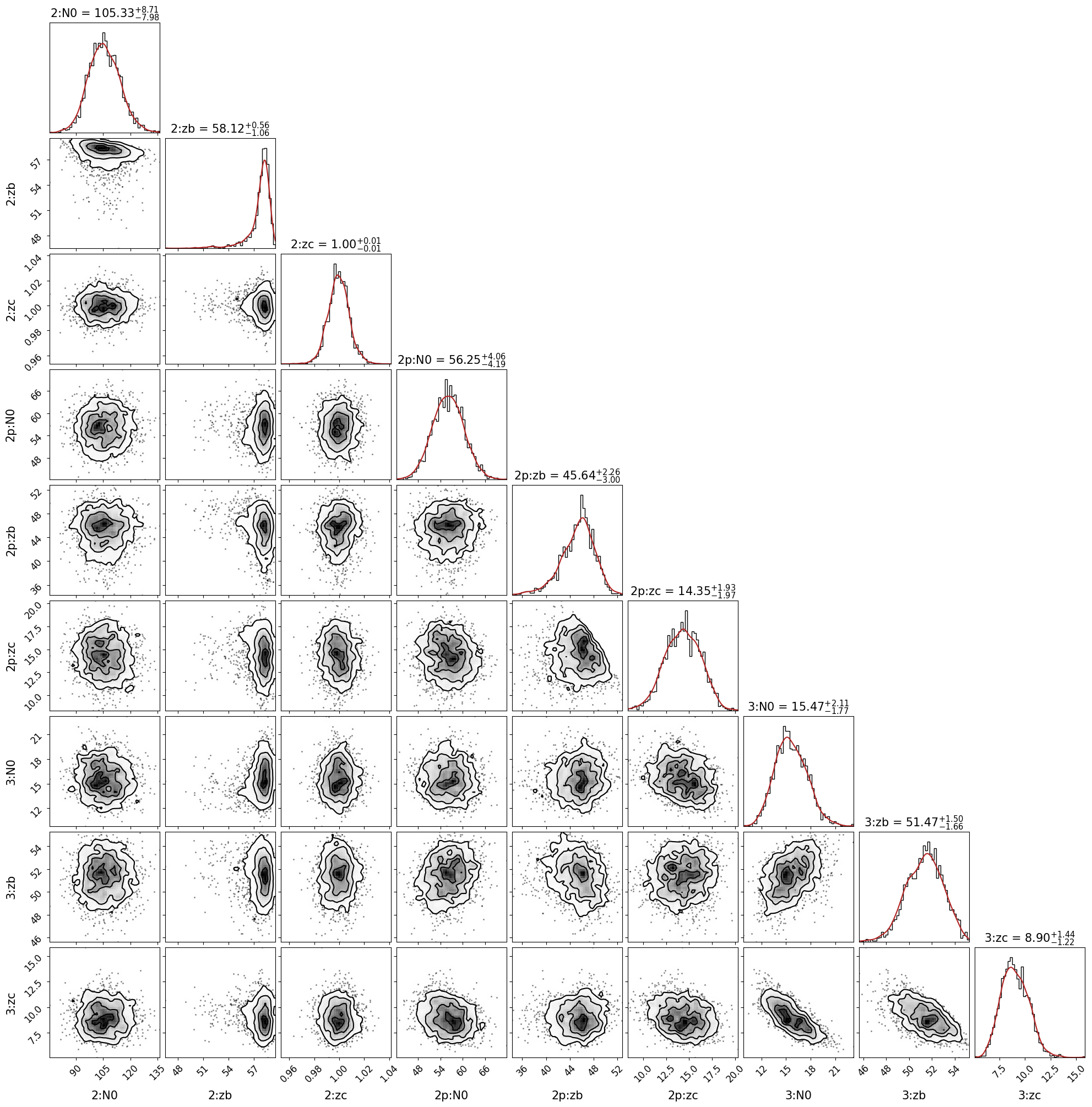}
    \caption{Corner plot of the retrieval for the `dark' cold collar location of data cube VI0025\_07. Marginalized posterior probability distribution functions of each parameter are shown in the upper panel of each column. The red lines show kernel density estimation smoothing of the functions with a Gaussian kernel. The rest of the plots show the marginalized posterior probability distributions as function of a pair of parameters. Units of each parameter are indicated in Table \ref{table:determination_parameters}.}
    \label{fig:corner_coldcollar_dark_0025_07}
\end{figure*}

\begin{figure*}
    \centering
    \includegraphics[width=1\linewidth]{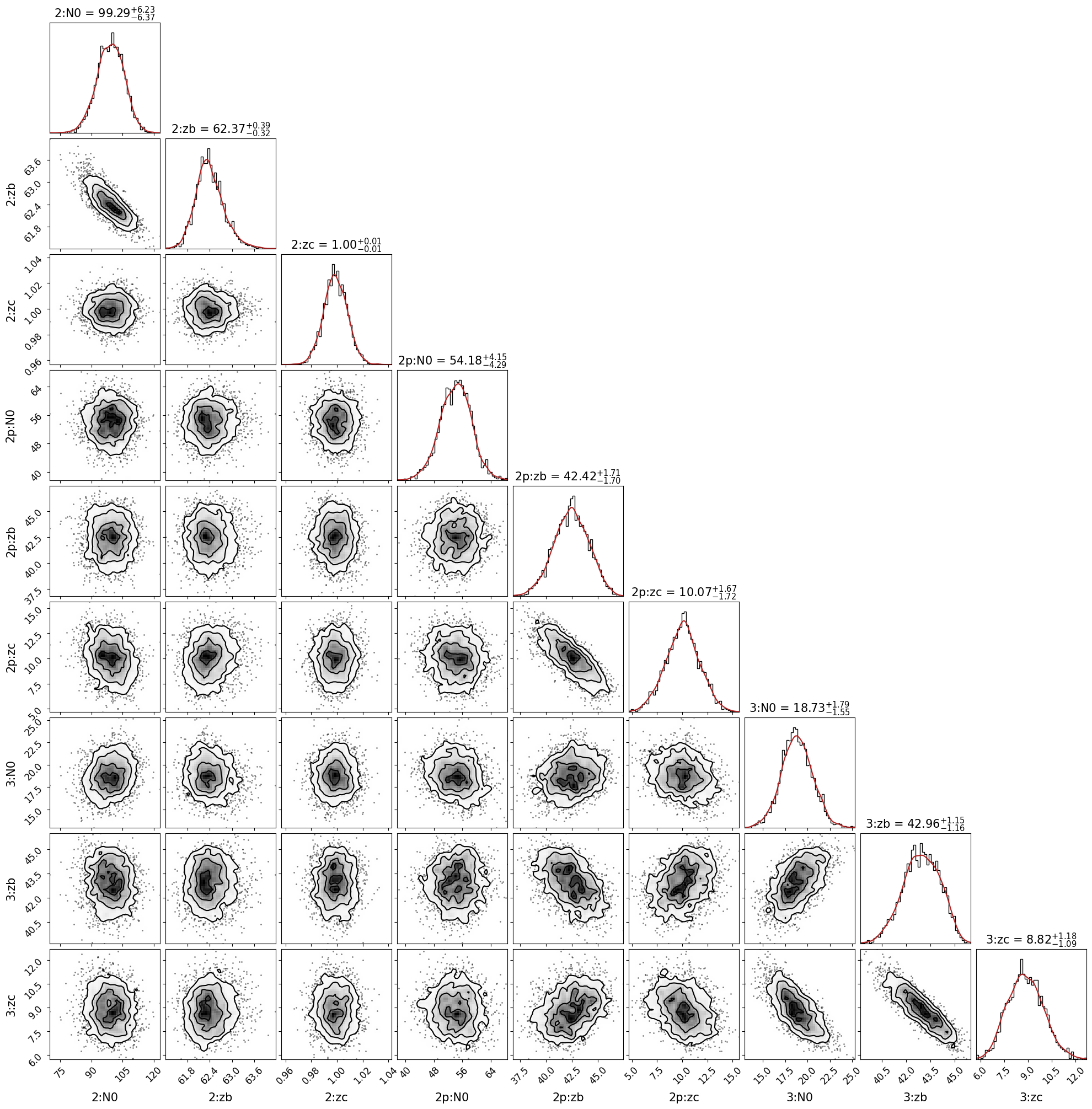}
    \caption{Corner plot of the retrieval for the `dark' South Polar Vortex location of data cube VI0025\_07. Marginalized posterior probability distribution functions of each parameter are shown in the upper panel of each column. The red lines show kernel density estimation smoothing of the functions with a Gaussian kernel. The rest of the plots show the marginalized posterior probability distributions as function of a pair of parameters. Units of each parameter are indicated in Table \ref{table:determination_parameters}.}
    \label{fig:corner_vortex_dark_0025_07}
\end{figure*}

\begin{figure*}
    \centering
    \includegraphics[width=1\linewidth]{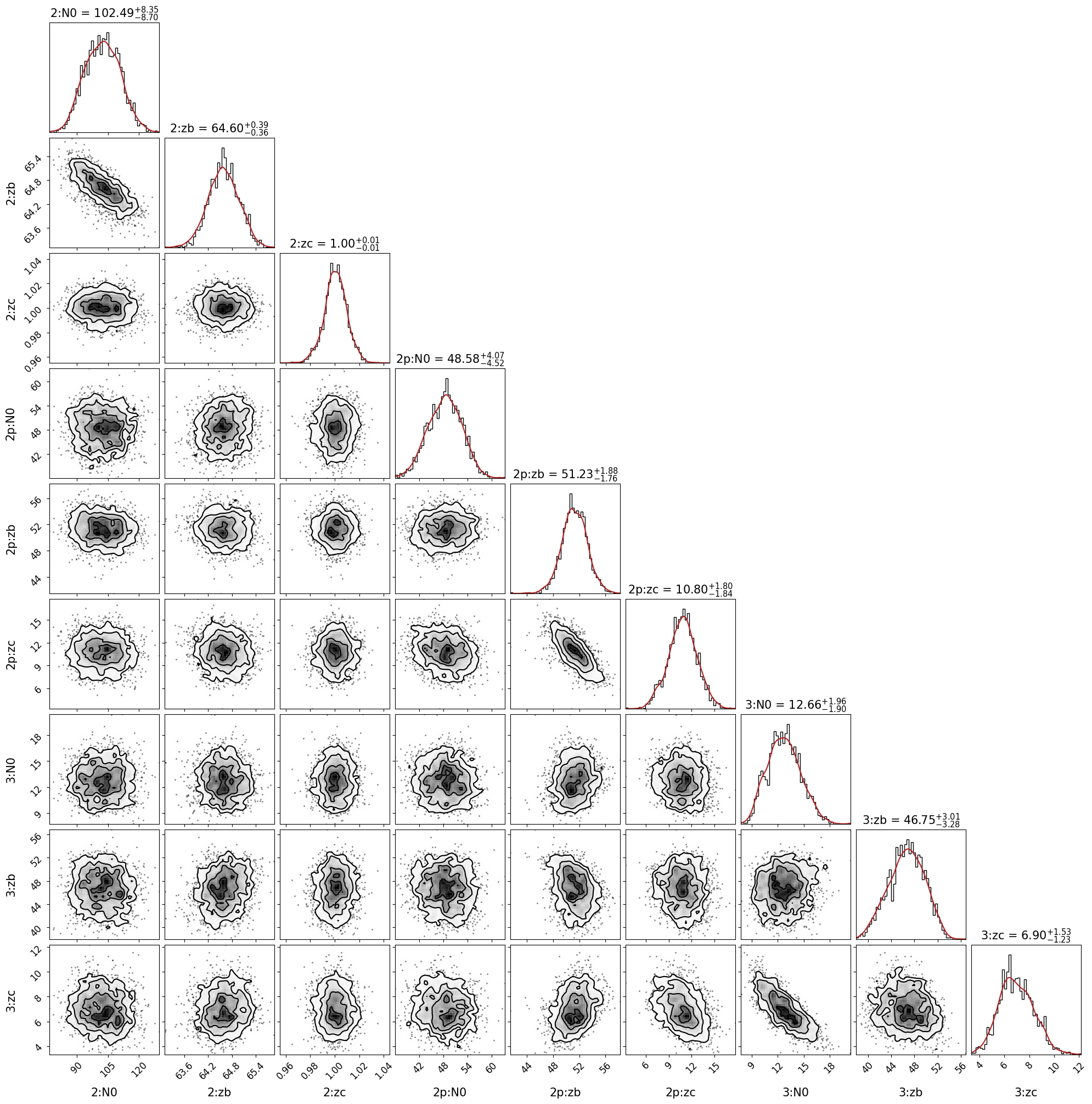}
    \caption{Corner plot of the retrieval for the `bright' mid-latitude location of data cube VI0818\_1. Marginalized posterior probability distribution functions of each parameter are shown in the upper panel of each column. The red lines show kernel density estimation smoothing of the functions with a Gaussian kernel. The rest of the plots show the marginalized posterior probability distributions as function of a pair of parameters. Units of each parameter are indicated in Table \ref{table:determination_parameters}.}
    \label{fig:corner_midlat_bright_0818_11}
\end{figure*}

\begin{figure*}
    \centering
    \includegraphics[width=1\linewidth]{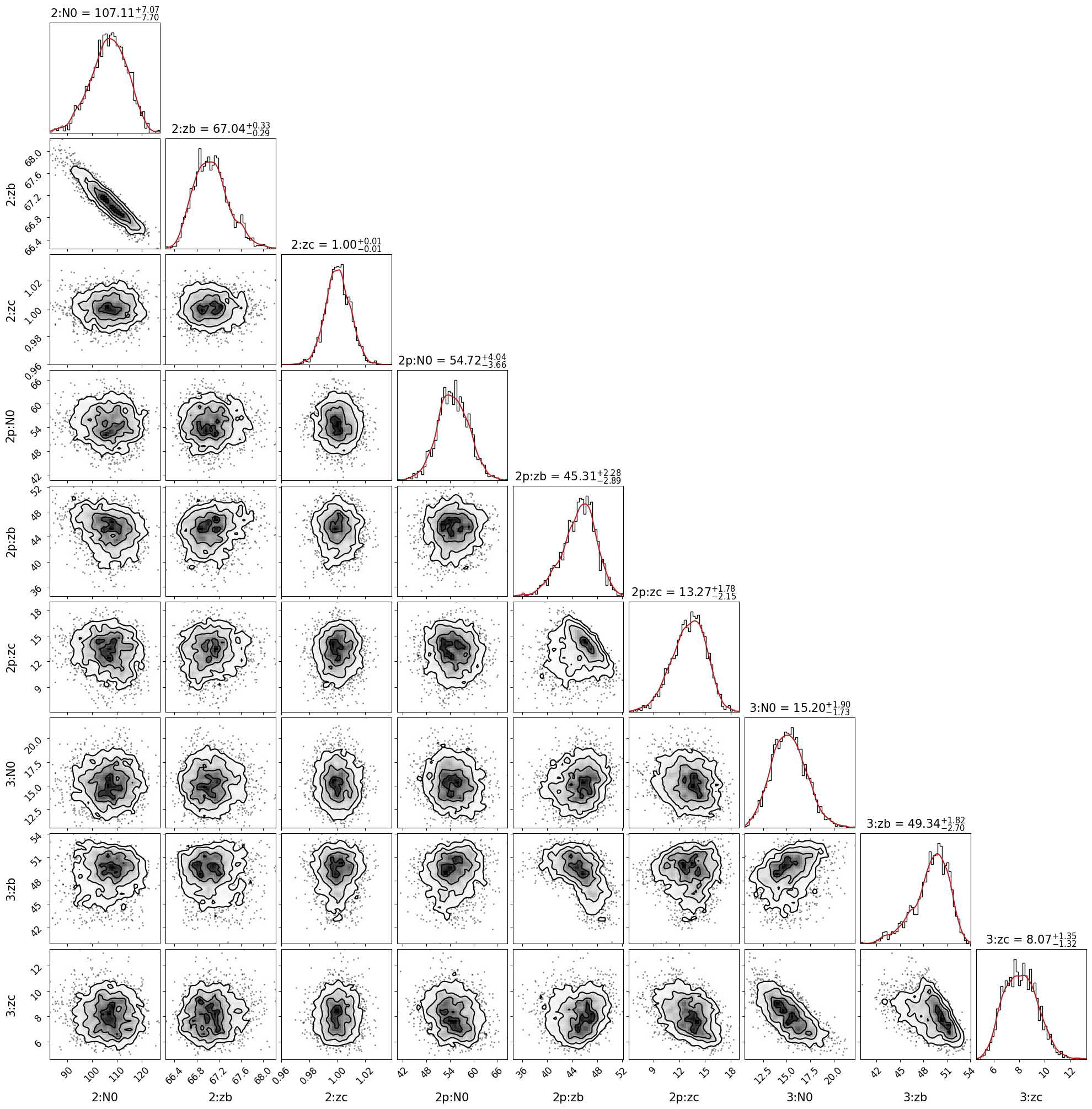}
    \caption{Corner plot of the retrieval for the `dark' mid-latitude location of data cube VI0411\_03. Marginalized posterior probability distribution functions of each parameter are shown in the upper panel of each column. The red lines show kernel density estimation smoothing of the functions with a Gaussian kernel. The rest of the plots show the marginalized posterior probability distributions as function of a pair of parameters. Units of each parameter are indicated in Table \ref{table:determination_parameters}.}
    \label{fig:corner_midlat_dark_0818_11}
\end{figure*}

\begin{figure*}
    \centering
    \includegraphics[width=1\linewidth]{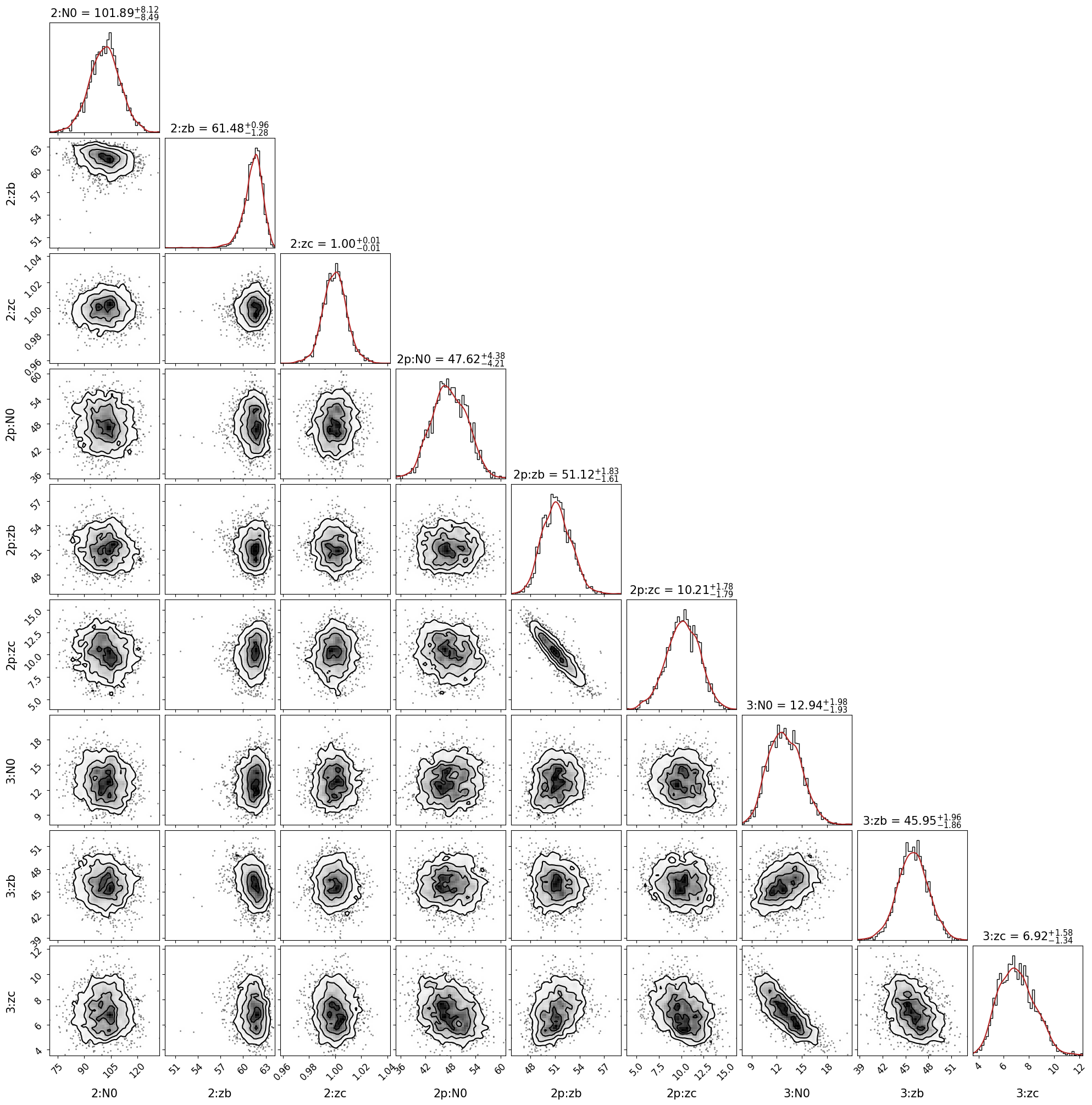}
    \caption{Corner plot of the retrieval for the `bright' cold collar location of data cube VI0726\_02. Marginalized posterior probability distribution functions of each parameter are shown in the upper panel of each column. The red lines show kernel density estimation smoothing of the functions with a Gaussian kernel. The rest of the plots show the marginalized posterior probability distributions as function of a pair of parameters. Units of each parameter are indicated in Table \ref{table:determination_parameters}.}
    \label{fig:corner_coldcollar_bright_0726_02}
\end{figure*}

\begin{figure*}
    \centering
    \includegraphics[width=1\linewidth]{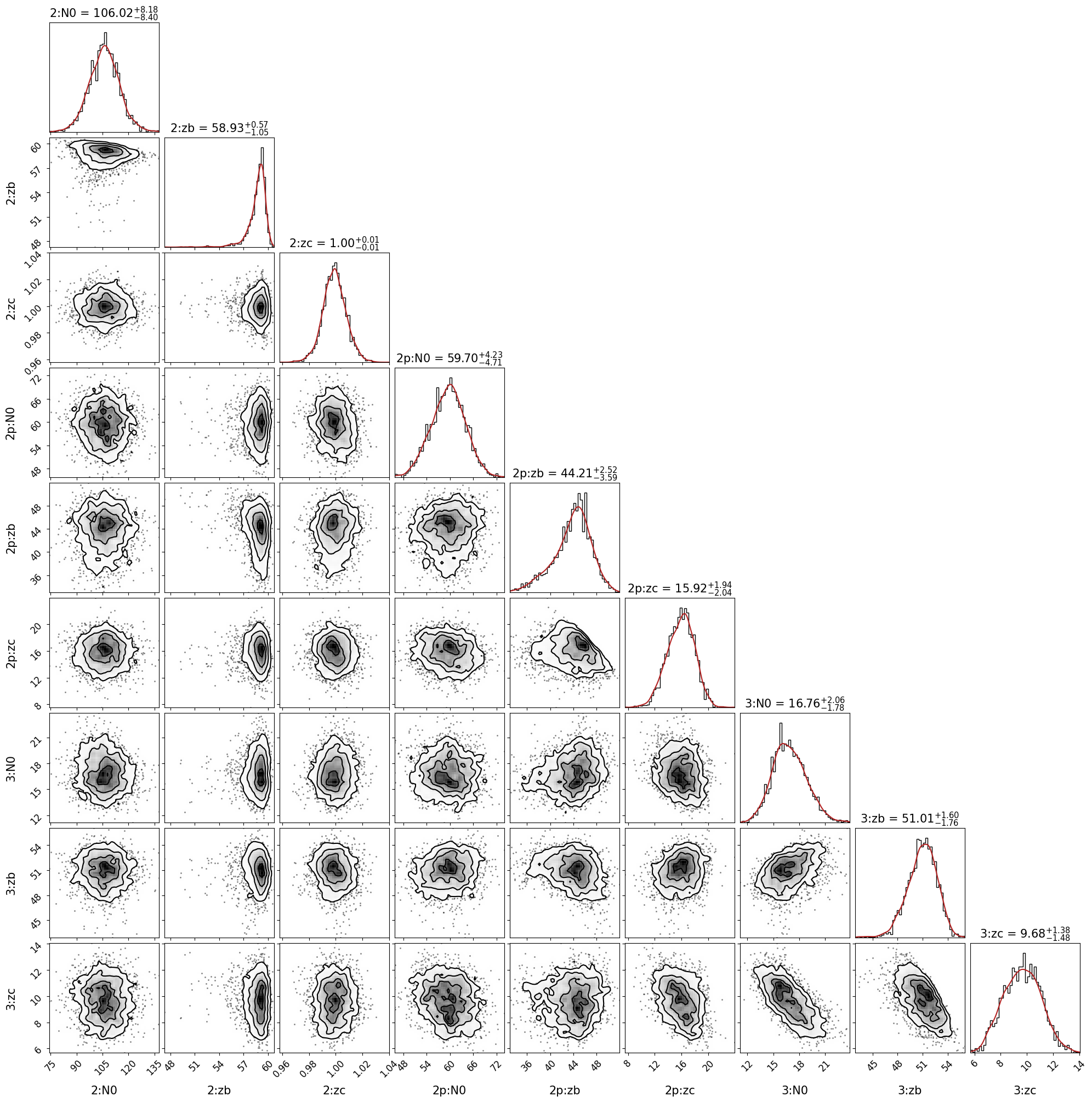}
    \caption{Corner plot of the retrieval for the `dark' cold collar location of data cube VI0095\_07. Marginalized posterior probability distribution functions of each parameter are shown in the upper panel of each column. The red lines show kernel density estimation smoothing of the functions with a Gaussian kernel. The rest of the plots show the marginalized posterior probability distributions as function of a pair of parameters. Units of each parameter are indicated in Table \ref{table:determination_parameters}.}
    \label{fig:corner_coldcollar_dark_0095_07}
\end{figure*}

\begin{figure*}
    \centering
    \includegraphics[width=1\linewidth]{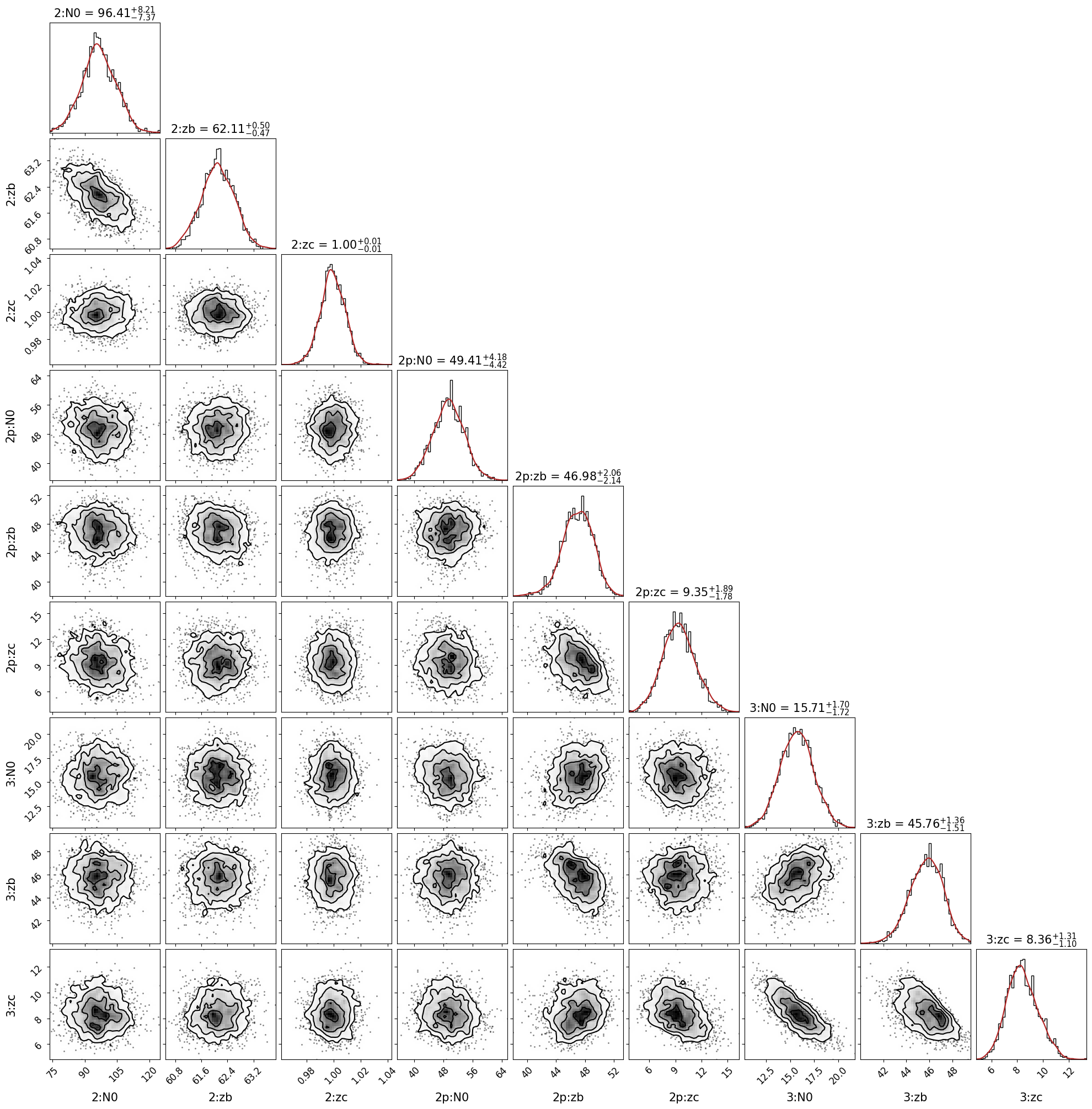}
    \caption{Corner plot of the retrieval for the `bright' South Polar Vortex location of data cube VI0038\_00. Marginalized posterior probability distribution functions of each parameter are shown in the upper panel of each column. The red lines show kernel density estimation smoothing of the functions with a Gaussian kernel. The rest of the plots show the marginalized posterior probability distributions as function of a pair of parameters. Units of each parameter are indicated in Table \ref{table:determination_parameters}.}
    \label{fig:corner_vortex_bright_0038_00}
\end{figure*}

\begin{figure*}
    \centering
    \includegraphics[width=1\linewidth]{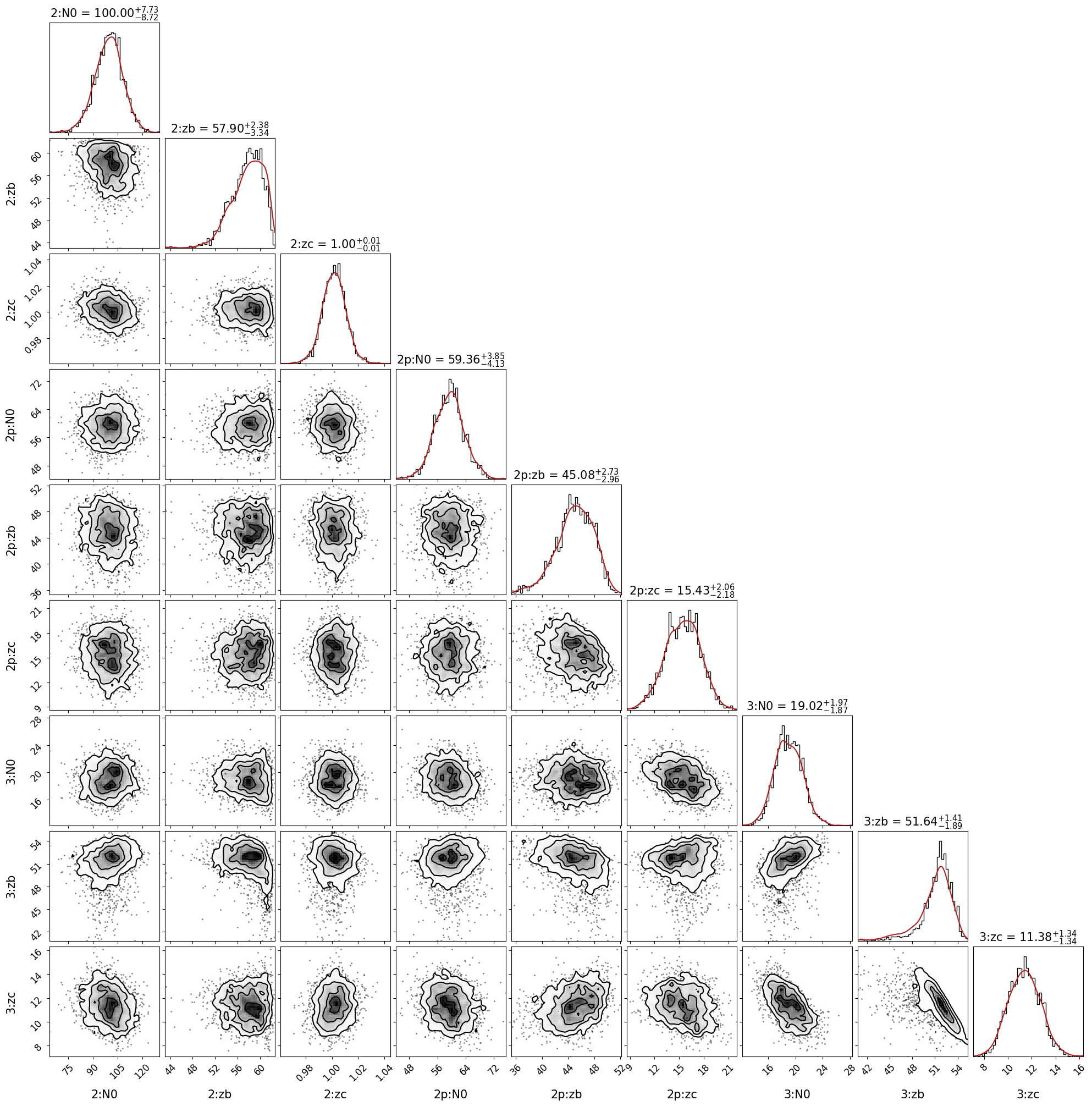}
    \caption{Corner plot of the retrieval for the `dark' South Polar Vortex location of data cube VI0310\_00. Marginalized posterior probability distribution functions of each parameter are shown in the upper panel of each column. The red lines show kernel density estimation smoothing of the functions with a Gaussian kernel. The rest of the plots show the marginalized posterior probability distributions as function of a pair of parameters. Units of each parameter are indicated in Table \ref{table:determination_parameters}.}
    \label{fig:corner_vortex_dark_0310_00}
\end{figure*}

\section{Spectral points} \label{sec:spectral_points}
The spectral points used in our retrievals are: 1.73, 1.74, 2.25, 2.29, 2.3, 2.32, 2.33, 2.34, 2.35, 2.39, 2.40, 3.80, 3.90, 4.00, 4.02, 4.05, 4.07, 4.08, 4.10, 4.11, 4.13, 4.15, 4.18, 4.47, 4.48, 4.49, 4.50, 4.51, 4.52, 4.53, 4.54, 4.55, 4.58, 4.6, 4.65, 4.67, 4.70, 4.75, 4.8, 4.82, 4.85, 4.90, 4.93, 4.95, 5.0, 5.02, 5.05, 5.07 and 5.10 $\mathrm{\mu}$m. The band-to-wavelength mapping of VIRTIS instrument depends on temperature \citep{kappel_refinements_2012}, and hence these values are approximated.

\end{document}